\documentclass[12pt]{article}
\usepackage{amsmath}
\usepackage{graphicx,psfrag,epsf}
\usepackage{enumerate}
\usepackage{natbib}
\usepackage{url} 
\graphicspath{{C:/Users/whsigris/Dropbox/HSLU/Projects/KTI_FraudDetection_CreditRisk/plots/Modeling/}{C:/Users/whsigris/Dropbox/HSLU/Projects/KTI_FraudDetection_CreditRisk/plots/}}
\usepackage{amssymb, amsmath, amsthm,graphicx,bbm,float}
\usepackage[linesnumbered,ruled]{algorithm2e}
\usepackage[toc,page]{appendix}

\DeclareMathOperator*{\argmin}{argmin}
\DeclareMathOperator*{\argmax}{argmax}

\newcommand{\blind}{0}

\begin{document}

\def\spacingset#1{\renewcommand{\baselinestretch}%
{#1}\small\normalsize} \spacingset{1}


\if0\blind
{
  \title{\bf Grabit: Gradient Tree-Boosted Tobit Models for Default Prediction}
  \author{Fabio Sigrist\thanks{Corresponding author. Email: fabio.sigrist@hslu.ch. Address: Institute of Financial Services Zug IFZ, Lucerne University of Applied Sciences and Arts, Grafenauweg 10, 6302 Zug, Switzerland.} \hspace{.2cm}\\
    Lucerne University of Applied Sciences and Arts\\
    and \\
    Christoph Hirnschall \\
    Advanon}
  \maketitle
}
\fi

\bigskip

\begin{abstract}
A frequent problem in binary classification is class imbalance between a minority and a majority class such as defaults and non-defaults in default prediction. In this article, we introduce a novel binary classification model, the \textbf{Grabit} model, which is obtained by applying \textbf{gra}dient tree \textbf{b}oosting to the Tob\textbf{it} model. We show how this model can leverage auxiliary data to obtain increased predictive accuracy for imbalanced data. We apply the Grabit model to predicting defaults on loans made to Swiss small and medium-sized enterprises (SME) and obtain a large and significant improvement in predictive performance compared to other state-of-the-art approaches. 

\end{abstract}

\noindent%

\vfill

\newpage
\spacingset{1.45} 

\bibliographystyle{abbrvnat}

\section{Introduction}\label{intro}
Prediction of corporate failures is important since bankruptcies can result in significant economic losses for investors and even cause economic downturns and recessions. A common problem in default prediction is that bankruptcies are rare events. This means that the number of defaulted companies is typically much lower than the number of non-defaulted ones. In general, class imbalance \citep{japkowicz2002class}, i.e., the situation where a majority class occurs much more frequently than a minority class, is a problem encountered in various other areas such as fraud detection, predictive maintenance, or generally the modeling of rare events such as stock market crashes or flood events. The problem of class imbalance is aggravated if the size of the dataset is small since in this case the available data for the minority class is even smaller.

In this article, we introduce a novel and flexible classification model, the Grabit model, which is obtained by applying \textbf{gra}dient tree \textbf{b}oosting to the Tob\textbf{it} model. The Grabit model allows for alleviating the class imbalance and small data problem and for obtaining increased predictive accuracy if data for an auxiliary variable, which is related to the underlying decision function\footnote{In the context of classification, the decision function denotes a function of the predictor variable that determines the decision boundary, i.e, the potentially nonlinear hyperplane in the space of the predictor variables which separates the different classes.}, is observed. We use the term auxiliary variable to denote a response variable that is observed for the majority class in addition to the binary default indicator. For instance for default prediction, this auxiliary variable can be in the form of stock returns, a distance-to-default measure, or credit spreads for companies with publicly traded stocks or bonds, or number of days of delay until repayment, number of outstanding or missing payments, or amount in arrears for non-public companies. Intuitively, the Grabit model can learn the relationship between the predictor variables and the binary response variable better when also using the auxiliary data. In our application, the auxiliary variable consists of days of delay until repayment and can be interpreted as a default potential. The idea is that companies that did not default but have a high default potential are similar to defaulted companies, and one can thus alleviate the class imbalance problem when applying a model that can use this information.


In our application, we show that the Grabit model considerably outperforms other state-of-the-art approaches for default prediction of loans made to Swiss small and medium-sized enterprises (SMEs). Further, in a simulation study, we investigate how the increased predictive accuracy of the Grabit model depends on (i) the dependence between the auxiliary variable and the latent decision function, (ii) the sample size, (iii) the class imbalance ratio, and (iv) the complexity of the decision function. We find that the larger the dependence between the auxiliary variable and the decision function, the smaller the sample size, and the larger the class imbalance, the higher is the accuracy gain of the Grabit model. Interestingly, if the auxiliary variable is independent of the decision function, i.e. the auxiliary data contains no additional information, the Grabit model still performs as well as the best competing binary classifier in our simulation study. Further, we observe that the Grabit model outperforms other models also in cases of larger datasets if the decision function is sufficiently complex, for example, having strong nonlinearities or interactions among predictors.

The Grabit model offers several important additional advantages over existing approaches. In contrast to linear Logit and Tobit models, the Grabit model can learn nonlinearities, discontinuities, and complex interactions. Since the Grabit model uses trees as base learners, it is robust against outliers in predictor variables and scale invariant to monotonic transformation for the predictor variables. This means that no transformation of the predictor variables is needed which is an important advantage in practice. Other advantages are that missing values in the predictors can be automatically accommodated and do not need to be imputed \citep{elith2008working} and that the predictive performance is not impaired by the problem of multicollinearity. 

The Grabit algorithm is implemented in the Python package \texttt{KTBoost}, which is openly available from the Python Package Index (PyPI) repository.\footnote{See https://github.com/fabsig/KTBoost for more information.}

\subsection{Prior literature on default prediction}
Default prediction has been of major interest to both researchers and practitioners in the financial sector for almost a century \citep{fitzpatrick1932comparison, winakor1935changes, merwin1942financing}. See \citet{altman2002bankruptcy} and \citet{bellovary2007review} for surveys on default prediction. In early proposals for statistical default prediction, the focus was primarily on linear classification models such as linear discriminant analysis (LDA) or logistic regression where the covariates enter the model in a linear combination \citep{altman1968financial,zmijewski1984methodological,lau1987five, shumway2001forecasting, altman2007modelling, ding2012class,bauer2014hazard,tian2015variable}. Recently, nonlinear methods such as generalized additive models, neural networks, classification trees, and ensemble methods have been proposed for default prediction. In particular, standard boosting algorithms have also been applied to default prediction \citep{alfaro2008bankruptcy,zikeba2016ensemble,xia2017boosted}. See \citet{brown2012experimental, lessmann2015benchmarking, jones2017predicting} for a comparison of machine learning based approaches for default prediction.

As the above list of references shows, default prediction is traditionally approached using binary classification models. However, as we show in this article, binary classifiers can have low predictive accuracy if the data is imbalanced and if the sample size is small. To cope with this, we introduce the Grabit model as a tool that allows for combining data for an auxiliary variable with the binary default data in order to obtain a default prediction model with higher predictive accuracy. Note that, in general, the use of some form of auxiliary data is not novel in default prediction as, for instance, multi-state models implicitly also use auxiliary data such as rating migrations. Examples of multi-state models include \citet{koopman2008multi} and \citet{djeundje2018incorporating}. Rating migrations can be modeled using binary classifiers \citep{jones2015empirical} or by using multiclass classifiers. Concerning the latter, see, e.g., \citet{vahid2016modeling} for a recent example of the application of a machine learning based multiclass classifier. In our application, we compare such a multi-state classifier to the Grabit model and observe that the Grabit model performs substantially and significantly better than the multi-state classifier.


\citet{moffatt2005hurdle} also applies the Tobit model and a double hurdle extension to default prediction with a different goal than ours, though. Conditional on that a default occurs, \citet{moffatt2005hurdle} also models the extent of default as measured by, e.g., the amount in arrears. This can then be used for a loss given default calculation.

The remainder of this article is structured as follows. In Section \ref{grabitmain}, we briefly review the Tobit model and gradient boosting and then introduce the Grabit model. In Section \ref{simstudy}, we investigate the performance of the Grabit model in a simulation study. We then apply the model to the prediction of defaults of Swiss small and medium-sized enterprises (SMEs) in Section \ref{defaulpred}. We conclude in Section \ref{concl} and mention possible directions for future research.

\section{Gradient Tree-Boosted Tobit Model}\label{grabitmain}
\subsection{The Tobit model}\label{tobitsec}
Auxiliary data for the majority class can be combined with the binary data by the use of censored regression models. In the following, we briefly present the two-sided version of the Tobit model \citep{tobin1958estimation, rosett1975estimation}, which is one of the most widely used censored regression models. The assumption of the Tobit model is that there exists a latent variable $Y^*$ which follows, conditional on some covariates $X = (X_1,\dots,X_p)^T\in\mathbb{R}^p$, a Gaussian distribution:
\begin{equation}\label{tobitmod1}
Y^*|X\sim N(F(X),\sigma^2).
\end{equation}
The mean $ F(X)$ is assumed to depend linearly on the covariates $X$ through $F(X)=X^T\beta$, where $\beta \in \mathbb{R}^p$ is a set of coefficients. This latent variable $Y^*$ is observed only if it lies in an interval $[y_l,y_u]$. Otherwise, one observes $y_l$ or $y_u$ depending on whether the latent variable is below the lower threshold $y_l$ or above the upper threshold $y_u$, respectively. Denoting $Y$ as the observed variable, we can express this as
\begin{equation*}
Y=\min(\max(Y^*,y_l),y_u).
\end{equation*}
Note that one-sided Tobit models are obtained as special cases by letting one of the boundaries $y_l$ or $y_u$ converge to minus or plus infinity. We refer to \citet{maddala1986limited} or \citet{amemiya1985advanced} for more information on the Tobit model.



Although the Tobit model is defined by a censoring mechanism, it can be applied not only to truly censored data but in many other situations where the data consists of a continuous part and a discrete point mass at the borders. This includes fractional response data \citep{papke1996econometric,papke2008panel}, loss given default \citep{moffatt2005hurdle, sigrist2011using}, rainfall \citep{sanso1999venezuelan, sigrist2012dynamic}, or default prediction as in this article. The latent variable $Y^*$ can often be interpreted as a potential which indicates how likely the event under consideration is to occur. In our case, $Y^*$ can be interpreted as a default potential, and a default occurs if the potential $Y^*$ exceeds a certain threshold. Default events thus correspond to the case $Y^*\geq y_u$, and the observed data is identified with $Y=y_u$. The non-default cases correspond to $Y^*< y_u$, and the auxiliary variable is identified with the observed variable $Y=Y^*$ in this case. 

In machine learning, the focus is usually not on econometric or statistical models but rather on loss functions which are minimized. Another justification for the use of the Tobit model is the fact that for the default cases $Y=y_u$, the Tobit loss, i.e., the negative log-likelihood, is asymmetric in the predictor function $F(X)$: the larger the default potential the lower is the loss, and the lower the default potential the larger is the loss. This is a desirable property since predicting a larger default potential above the default threshold in case a default occurs is indeed preferable, whereas predicting a low default potential is undesirable in this case. Or, in other words, a symmetric loss functions such as the squared loss is clearly not desirable since a high predicted default potential should not result in a larger loss for default cases. From this point of view, the Tobit can thus also be simply thought of as a way of obtaining an appropriate asymmetric loss function. 

Further, we do not necessarily assume that the observed default data and the auxiliary data is generated by Tobit model. We rather consider the Tobit likelihood as a tool that allows for combining the binary default data with the auxiliary data. As we show in the simulation study in Section \ref{simstudy}, gains in predictive accuracy can also be obtained with the Grabit model in cases where the auxiliary data and the binary default data are simulated by different data generating processes.


\subsection{Boosting}\label{gradboost}
A rather restrictive assumption of the Tobit model is the linear function which relates a set of covariates to a linear predictor. In this article, we relax this assumption by applying \textbf{gra}dient tree \textbf{b}oosting to the Tob\textbf{it} model. We denote the resulting model as 'Grabit' model. Boosting enjoys large popularity in many areas mainly due to its high predictive accuracy on a wide range of datasets; see, e.g., \citet{chen2016xgboost} or \citet{yang2017insurance}. It is an ensemble technique which additively combines multiple relatively simple models, so-called base learners which often consist of regression trees. Boosting was first introduced in machine learning for classification \citep{freund1995desicion}. Important contributions to the topic, in particular, the statistical view of boosting as stagewise optimization of a risk functional include \citet{breiman1998arcing,friedman2000additive,friedman2001greedy}. See \citet{buhlmann2007boosting}, \citet{mayr2014evolution}, and \citet{sigrist2018gradient} for reviews and overviews on boosting algorithms. 


In the following, we briefly present the idea of boosting approach as introduced by \citet{friedman2001greedy}. We assume that there is a response variable $Y$ and a vector of covariates $X\in \mathbb{R}^p$, and that we observe data $(y_i,x_i), i=1,\dots,n$. The goal of boosting is to find a minimizer $F^*(\cdot)$ of the empirical loss\footnote{In machine learning and statistics, this is usually called 'empirical risk'. We use the term 'empirical loss' instead of 'empirical risk' to avoid confusion with the term risk in a finance context.} $R^e(F)$
\begin{equation}\label{emprisk}
\begin{split}
F^*(\cdot)&=\argmin_{F(\cdot)\in \Omega_{\mathcal{S}}}R^e(F)\\
&=\argmin_{F(\cdot)\in \Omega_{\mathcal{S}}}\sum_{i=1}^n L(y_i,F(x_i)),
\end{split}
\end{equation}
where $F(\cdot): \mathbb{R}^p\rightarrow \mathbb{R}$ are functions that map $X$ to $Y$ and $L$ is an appropriately chosen loss function such as, e.g., the squared loss $L(y,F)=(y-F)^2/2$ or the negative Tobit log-likelihood in our case. In boosting, one restricts the functions $F(\cdot)$ to lie in the span $\Omega_{\mathcal{S}}=span(\mathcal{S})$ of a set $\mathcal{S}$ of so-called base learners $h\left(x;a^{[m]}\right)$. Specifically, boosting constructs an ensemble
$$F(x)=F^{[0]}+\sum_{m=1}^M\rho^{[m]}h\left(x;a^{[m]}\right),$$
where we assume that $h\left(x;a^{[m]}\right)$ are regression trees \citep{breiman1984classification} with parameters $a^{[m]}$, $F^{[0]}$ is a constant, $\rho^{[m]}\in \mathbb{R}$, and $M$ denotes the number of boosting iterations or trees.

The boosting approach of \citet{friedman2001greedy} iteratively finds $F^*(\cdot)$ using a functional gradient descent algorithm. Denoting the current estimate for $F^*(\cdot)$ by $F^{[m-1]}(\cdot)$, an update from $F^{[m-1]}(\cdot)$ to $F^{[m]}(\cdot)$ is obtained by first calculating the negative gradient $-\frac{\partial L(y_i,F(x_i))}{\partial F}\Big|_{F=F^{[m-1]}}$, and then approximating this gradient with a base learner $h\left(x;a^{[m]}\right)$. If trees are used as base learners and the second derivative of the loss function $L(y,F)$ exists, is non-constant and non-zero, \citet{friedman2001greedy} suggests to do an additional step of Newton's method to find the leaf values. One thus performs a hybrid gradient-Newton approach where the partition of the trees are learned using gradient descent and the leaf values are learned using Newton's method. See \citet{sigrist2018gradient} and Section \ref{grabit} for more details. 


In addition, a shrinkage factor $\nu$, $0<\nu\leq 1$ is typically used for the update step:
$$F^{[m]}(x)=F^{[m-1]}(x)+\nu\rho^{[m]}h\left(x;a^{[m]}\right).$$ 
This parameter $\nu$ acts as a regularization parameter. It has been empirically observed that the introduction of a shrinkage factor slows down overfitting and results into increased predictive performance \citep{friedman2001greedy}.

\subsection{The Grabit model}\label{grabit}
In this section, we introduce the Grabit model, which is obtained by extending the Tobit model using gradient boosting with trees as base learners.  While nonlinearities and interactions can be modeled in various ways, boosting with trees provides a very flexible approach that relies on few assumptions and which, in particular, shows very good predictive performance on a wide range of datasets \citep{chen2016xgboost}.

Instead of assuming a linear function for the mean function $F(X)$ of the latent variable $Y^*$, the Grabit model uses a flexible function ${F}(\cdot): \mathbb{R}^p\rightarrow \mathbb{R}$ which consists of an ensemble of regression trees. An estimate for this function is found by applying boosting with regression trees as base learners. Specifically, we use the negative log-likelihood of the Tobit model as loss function $L(y,F)$:
$$
L(y,F)=-\log\left(f_{F,\sigma}(y)\right),
$$
where both the density $f_{F,\sigma}(y)$ of the Tobit model and the corresponding loss $L(y,F(x))$ are given in Equations \eqref{tobitdensity} and \eqref{tobitloss} in Appendix \ref{Tdensity}. 

The boosting approach of \citet{friedman2001greedy} then works by iteratively fitting a regression tree $h(x_i,a^{[m]})$ as a least squares approximation to the so-called pseudoresponses $\widetilde{y}_i$ which equal the negative gradient 
$$ \widetilde{y}_i=-\frac{\partial L(y_i,F)}{\partial F}\Big|_{F=F^{[m-1]}(x_i)}.$$ 
One thus obtains a $J$ terminal node tree $h(x_i,a^{[m]})$ whose partition of the space we denote by $\{R_{j}^{[m]}\}_{j=1}^J$. Next, the optimal $\rho^{[m]}$ is found by minimizing the empirical loss
$$ \rho^{[m]}=\argmin_{\rho}\sum_{i=1}^nL(y_i,F^{[m-1]}(x_i)+\rho h(x_i,a^{[m]})).$$ 
If trees are used as base learners, instead of finding one global step size $\rho^{[m]}$, we can find the step size for each terminal node $R_{j}^{[m]}$ separately by recalculating the optimal values of the terminal nodes $\gamma_{j}^{[m]}$:
$$\gamma_{j}^{[m]}=\displaystyle \argmin_{\gamma}\sum_{x_i\in R_{j}^{[m]}}L(y_i,F^{[m-1]}(x_i)+\gamma).$$
Since for the Tobit loss function this line search cannot be done in closed form, an approximate minimization method has to be used. Following \citet{friedman2001greedy}, we use a second order Taylor approximation for $\sum_{x_i\in R_{j}^{[m]}}L(y_i,F^{[m-1]}(x_i)+\gamma)$ around $F^{[m-1]}(x)$, and find $\gamma_{j}^{[m]}$ such that this approximation is minimized. This corresponds to performing a single Newton-Raphson step as follows:
\begin{equation*}\label{coefpart}
\gamma_{j}^{[m]}=-\sum_{x_i\in R_{j}^{[m]}}\frac{\partial L(y_i,F)}{\partial F}\Big|_{F=F^{[m-1]}(x_i)}\Bigg/\sum_{x_i\in R_{j}^{[m]}}\frac{\partial^2 L(y_i,F)}{\partial^2 F}\Big|_{F=F^{[m-1]}(x_i)}.
\end{equation*}

For the boosting algorithm described above, we need to be able to evaluate both the gradient $\frac{\partial L(y_i,F)}{\partial F}$ and the second derivative $\frac{\partial^2 L(y_i,F)}{\partial^2 F}$. These can be calculated explicitly; see Equations \eqref{gradient} and \eqref{secderiv} in Appendix \ref{Tdensity}. In summary, we use a gradient descent step to find the structure of the trees, i.e., the partition of the space, and a Newton update step to learn the leaf values. Algorithm \ref{alg:algorithm-label} summarizes this. Note that we consider $\sigma$ as a known parameter. The parameter $\sigma$ can be chosen by cross-validation or estimated using a profile-likelihood approach; see Section \ref{tunepar}.

\begin{algorithm}[ht!]
	\label{alg:algorithm-label}
	\caption{{\bfseries Grabit} ({\bfseries gra}dient tree-{\bfseries b}oosted Tob{\bfseries it} model)}
	Initialize $F^{[0]}(x)=\bar{y}.$\\
	\For{$m=1$ \KwTo $M$}{
		Compute the pseudoresponses using Equation \eqref{gradient}: $\widetilde{y}_i=-\frac{\partial L(y_i,F)}{\partial F}\Big|_{F=F^{[m-1]}(x_i)}, i=1\dots n$.\\
		Fit a $J$-terminal node regression tree to $(\widetilde{y}_i,x_i)$ using least squares and obtain a partition $R_{j}^{[m]},j=1,\dots J$.\\
		Update the terminal nodes of the tree using Equation \eqref{secderiv}: $\gamma_{j}^{[m]}=\sum_{x_i\in R_{j}^{[m]}}\widetilde{y}_i\Bigg/\sum_{x_i\in R_{j}^{[m]}}\frac{\partial^2 L(y_i,F)}{\partial^2 F}\Big|_{F=F^{[m-1]}(x_i)}.$\\
		Update $F^{[m]}(x)$: 
		$ F^{[m]}(x) = F^{[m-1]}(x)+\nu \sum_{j=1}^J\gamma_{j}^{[m]}\mathbbm{1}_{R_{j}^{[m]}}(x). $
	}
	Return $F^{[M]}(x)$.
\end{algorithm}

The Grabit algorithm is implemented in the Python package \verb|KTBoost|, which is openly available in the Python Package Index (PyPI) repository.\footnote{See https://github.com/fabsig/KTBoost for more information.}

\subsection{Choice of tuning parameters}\label{tunepar}
The Grabit algorithm has several tuning parameters which include the number of trees $M$, the shrinkage factor $\nu$, and the depth of the trees $T$. In addition, the standard deviation $\sigma$ of the latent variable $Y^*$ in Equation \eqref{tobitmod1} also needs to be estimated. In the following, we discuss how these parameters can be chosen.

The shrinkage factor $\nu$ and the number of trees $M$ control the amount of regularization. Past research \citep{friedman2001greedy, buhlmann2007boosting} has shown that the predictive accuracy of boosting algorithms is generally superior when choosing smaller values for $\nu$. The depth of the trees $T$ controls the degree of interaction among the covariates $X$. A tree of depth $T$ can maximally have interactions of order $T-1$. The parameters $\nu$, $M$, and $T$ can be chosen by cross-validation \citep[see, e.g.,][]{friedman2001elements} or using an information criterion. 


The parameter $\sigma$ can be chosen by either maximizing the profile likelihood or also by cross-validation. The profile log-likelihood function for $\sigma$ is given by the negative empirical loss $$\ell(\sigma)=-R^e(\widehat{F}_{\sigma},\sigma)$$ as a function of $\sigma$, where $\widehat{F}_{\sigma}(\cdot)$ is obtained as outlined above in Section \ref{grabit} for a fixed $\sigma$. The maximum
$$\widehat\sigma=\argmax_{\sigma}\ell(\sigma)$$
can be found by using a general purpose optimizer in the form of, e.g., a quasi-Newton method such as the Broyden-Fletcher-Goldfarb-Shanno algorithm. In order to avoid problems with negative values, one can reparametrize $\sigma$ by $\phi=\log(\sigma)\in \mathbb{R}$, find
$\widehat\phi=\argmax_{\phi}\ell(e^\phi),$
and set 
$\widehat\sigma=e^{\widehat\phi}.$
A computationally faster but potentially less accurate alternative is to do a grid search over a grid $\{\phi_1,\dots,\phi_K\}$, where $K$ is the number of grid points.

\section{Simulation study: class imbalance with auxiliary data}\label{simstudy}
In the following, we compare the Grabit model with other state-of-the-art approaches in a simulation study. We consider the task of binary classification in a class imbalance setting, where the minority class has relatively few observations, but auxiliary data correlated with the decision function is available for the majority class. The setting is inspired by the default prediction application in Section \ref{defaulpred}, where default cases are rare but one has auxiliary information for the non-default cases such as the number of days in delay, stock price returns, or distance-to-default measures. 

Overall, the goal of the simulation study is to show that whether the Grabit model provides increased predictive accuracy depends on (i) the dependence between the auxiliary variable and the latent decision function, (ii) the sample size, (iii) the class imbalance ratio, and (iv) also the complexity of the decision function.


\subsection{Dependence between the auxiliary data and the decision function}\label{simcor}
In the following, we consider four different levels of (Pearson) correlation, $0.75,0.5,0.25,$ and $0$, between the latent decision function and the auxiliary variable. We use a simulation setting that approximately mimics the situation of our application in Section \ref{defaulpred} in terms of the class imbalance ratio, the sample size $n$, and the number of predictor variables $p$. 

We assume the following data generating process:
\begin{equation}\label{classsim}
\begin{split}
Y^*&=F(X)+\epsilon, ~~X=(X_1,\dots,X_p)^T\in \mathbb{R}^p,~~\epsilon \sim N(0,\sigma_{\epsilon}^2),\\
C&=\mathbbm{1}_{\{Y^* \geq y_u\}}(Y^*),\\
Y_{a}&=C\cdot y_u+(1-C)\cdot\left(F(X)+\epsilon_a\right), ~~\epsilon_a \sim N(\mu_a,\sigma_a^2).
\end{split}
\end{equation}
Here, $C$ denotes the observed binary variable. The variable $C$ equals one if the latent variable $Y^*$ is above a threshold $y_u$ and zero otherwise. We assume that besides the binary variable $C$, we additionally observe the auxiliary variable $Y_a$. For the majority class $C=0$, the variable $Y_a$ equals the decision function $F$ plus a Gaussian error term $\epsilon_a$. This means that for the majority class, we have auxiliary data, which is correlated with the latent decision function $F$, and the standard deviation $\sigma_a$ of the noise term $\epsilon_a$ determines the correlation between $Y_a$ and $F$.

Simulation is done using the following specifications. For the mean function $F(\cdot)$ of the latent variable $Y^*$, we use $p=30$ and a nonlinear function of the following form:
\begin{equation}\label{decfct}
F(X)=\sum_{k=1}^{5} 0.3(X_{k})_+ +\sum_{k=1}^{3}\sum_{j=k+1}^{4} (X_{k}X_{j})_{+},~~X_k \overset{\text{ iid }}{\sim} \text{Unif}(-1,1),
\end{equation}
where $(x)_+=\max(x,0)$. With this choice, the first-order terms are nonlinear but monotone, the second-order interactions are also nonlinear, and only the first five variables of $X$ have a non-zero impact.\footnote{Note that the specific functional form used is not crucial for the results shown in the following, and very similar results are found using other nonlinear functional forms such as, e.g., simply using a sum of squares $F(X)=\sum_{k=1}^{p} X_{k}^2$ (results not tabulated). See also Section \ref{decfunc} for other choices of $F$.} The threshold $y_u$ is chosen such that we obtain a class imbalance ratio of approximately 95\% to 5\%, i.e., $P(C=1)\approx5\%$. For the above specified mean function, this corresponds to $y_u=2.84$. Further, we choose different values for $\sigma_a$ in order to compare the performance of the Grabit model under different correlation levels. Specifically, we use $\sigma_a=0.5, 0.98$, and $2.2$, which correspond to Pearson correlations between $Y_a$ and $F$ of approximately $0.75, 0.5$, and $0.25$. The means $\mu_a$ of $\epsilon_a$ are simply set to a value such that the auxiliary data is below the threshold $y_u$.\footnote{We use $\mu_a=-4,-5$, and $-9$.} In addition, we also consider the case of zero correlation between the auxiliary variable and the decision function by simulating according to $Y_{a}=C\cdot y_u+(1-C)\cdot\epsilon_a$, $\epsilon_a \sim N(-4,1)$. Finally, the standard deviation $\sigma_{\epsilon}$ of the noise term $\epsilon$ is chosen such that we have a signal-to-noise ratio of approximately one.\footnote{With the above specifications, this yields $\sigma_{\epsilon}=0.7$.}

We compare the one-sided Grabit model with two other classification methods which only use data for the binary variable $C$ and with the Tobit model which also uses the auxiliary data. As classification methods, we use a logistic regression model denoted as 'Logit' model as well as a tree-boosted classifier for a Bernoulli likelihood with a logistic link function denoted as 'boosted Logit' model.\footnote{For simplicity, we restrict ourselves to these two classifiers. However, other state-of-the-art classifiers such as random forest do not perform better than the boosted Logit model in our simulations (results not tabulated).} All computations are done in Python. The Logit and boosted Logit models are fitted using the Python package \texttt{scikit-learn} \citep{scikit-learn}.  For estimating the Grabit model, we use the algorithm presented in Section \ref{grabit} and implemented in the Python package \texttt{KTBoost}. For the Tobit model, we numerically minimize the negative log-likelihood in Equation \eqref{tobitloss} using a quasi-Newton method.

In each simulation iteration, we simulate $n=500$ data points as training data for estimating the models and additional $500$ data points as test data for comparing the different models. In total, the simulations are repeated $100$ times. Tuning parameters are selected on an additional independent validation dataset of the same size. We use the area under the receiver operating characteristic (AUROC) as a measure of the predictive accuracy for choosing tuning parameters. For the boosted Logit model, we consider the following tuning parameters: the number of trees $M$, the learning rate $\nu$, and the depth of the trees $T$. These are chosen among the following combinations of tuning parameters $M\in\{10,100,1000\}$, $\nu\in\{0.1,0.01,0.001\}$, and $T\in\{3,5,10\}$. For the Grabit model, we additionally select $\sigma$ from $\{0.01,0.1,1,10,100\}$. 

\begin{figure}[ht!]
	\centering
	\includegraphics[width=0.49\textwidth,trim={1.25cm 1.5cm 1.25cm 1.5cm},clip=true]{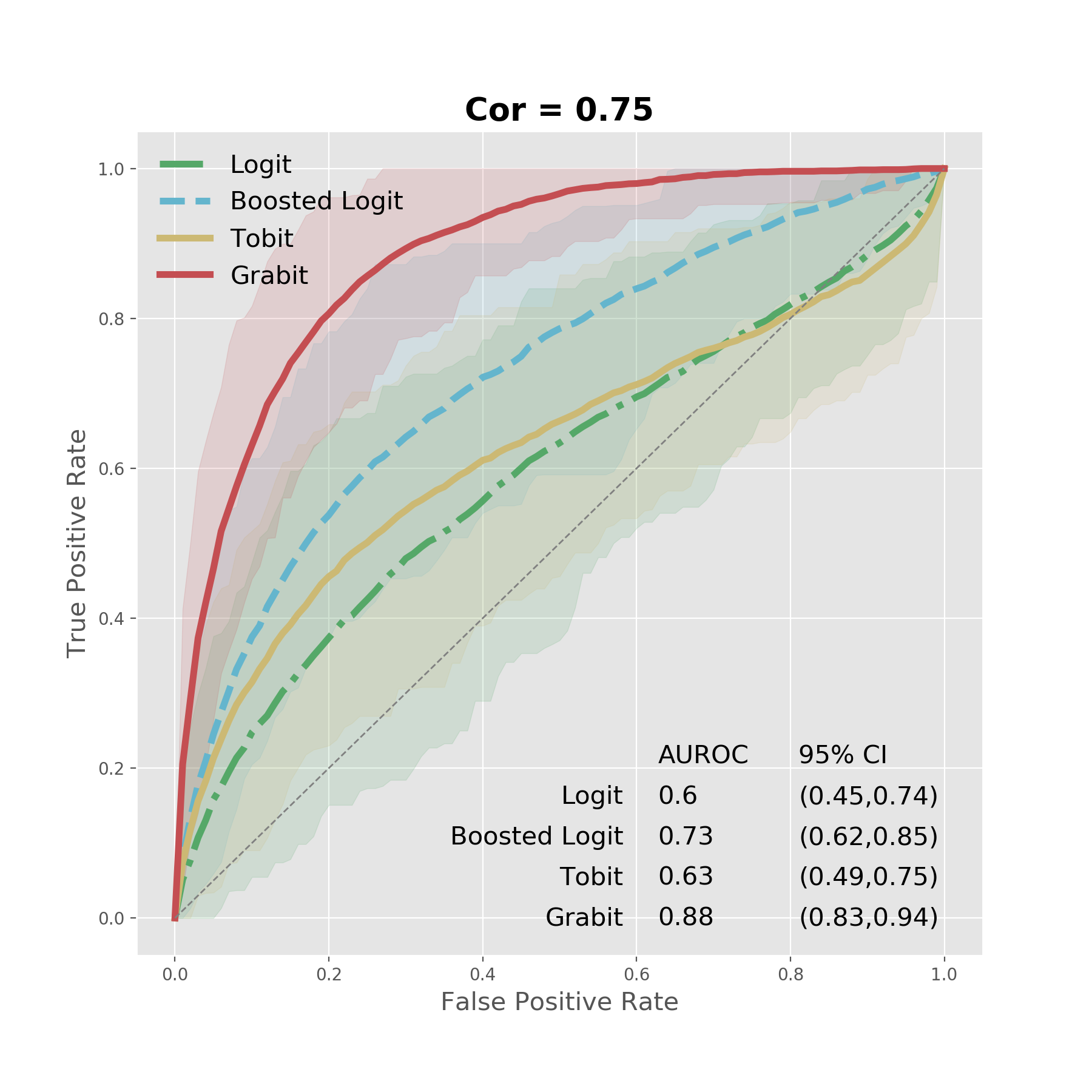}
	\includegraphics[width=0.49\textwidth,trim={1.25cm 1.5cm 1.25cm 1.5cm},clip=true]{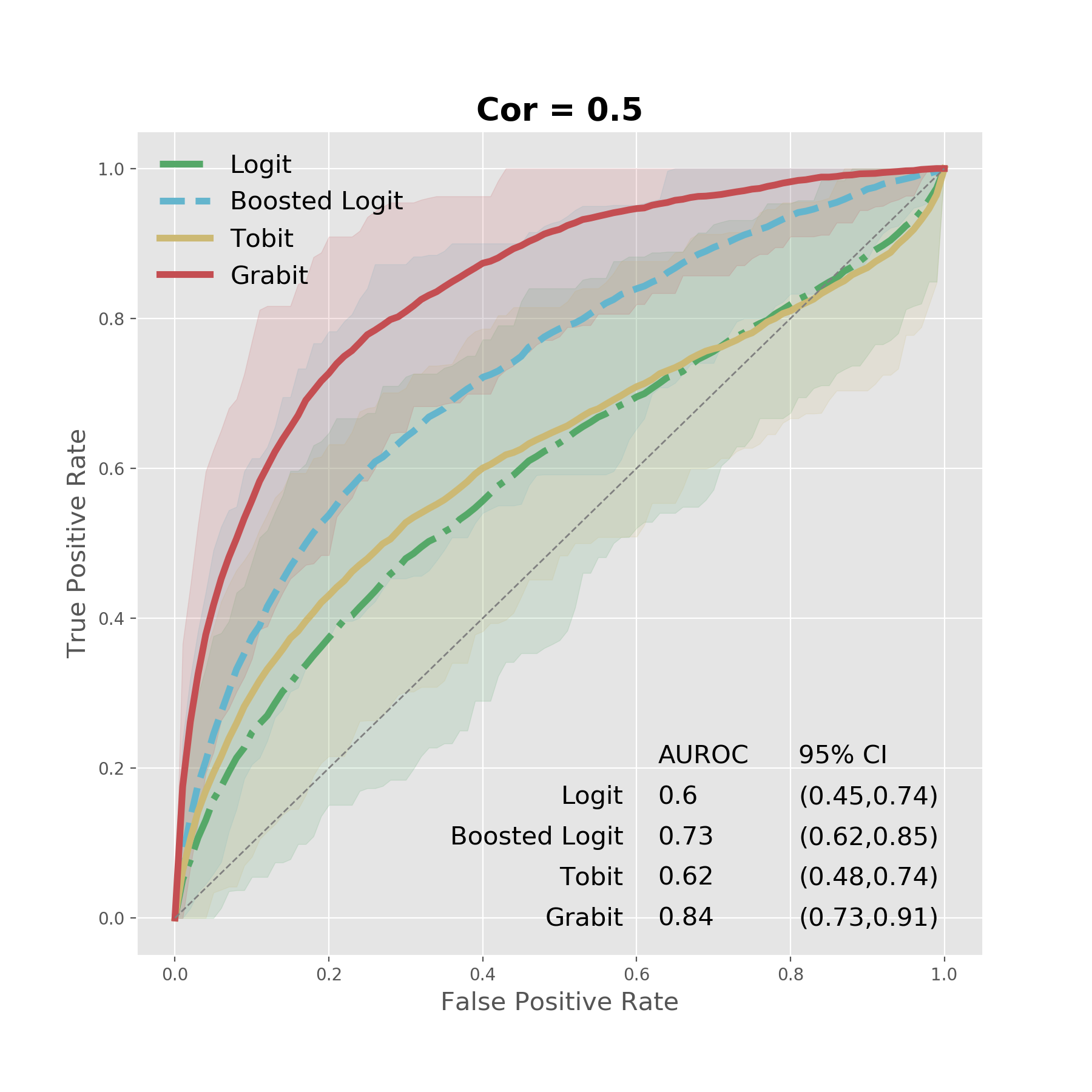}
	\includegraphics[width=0.49\textwidth,trim={1.25cm 1.5cm 1.25cm 1.5cm},clip=true]{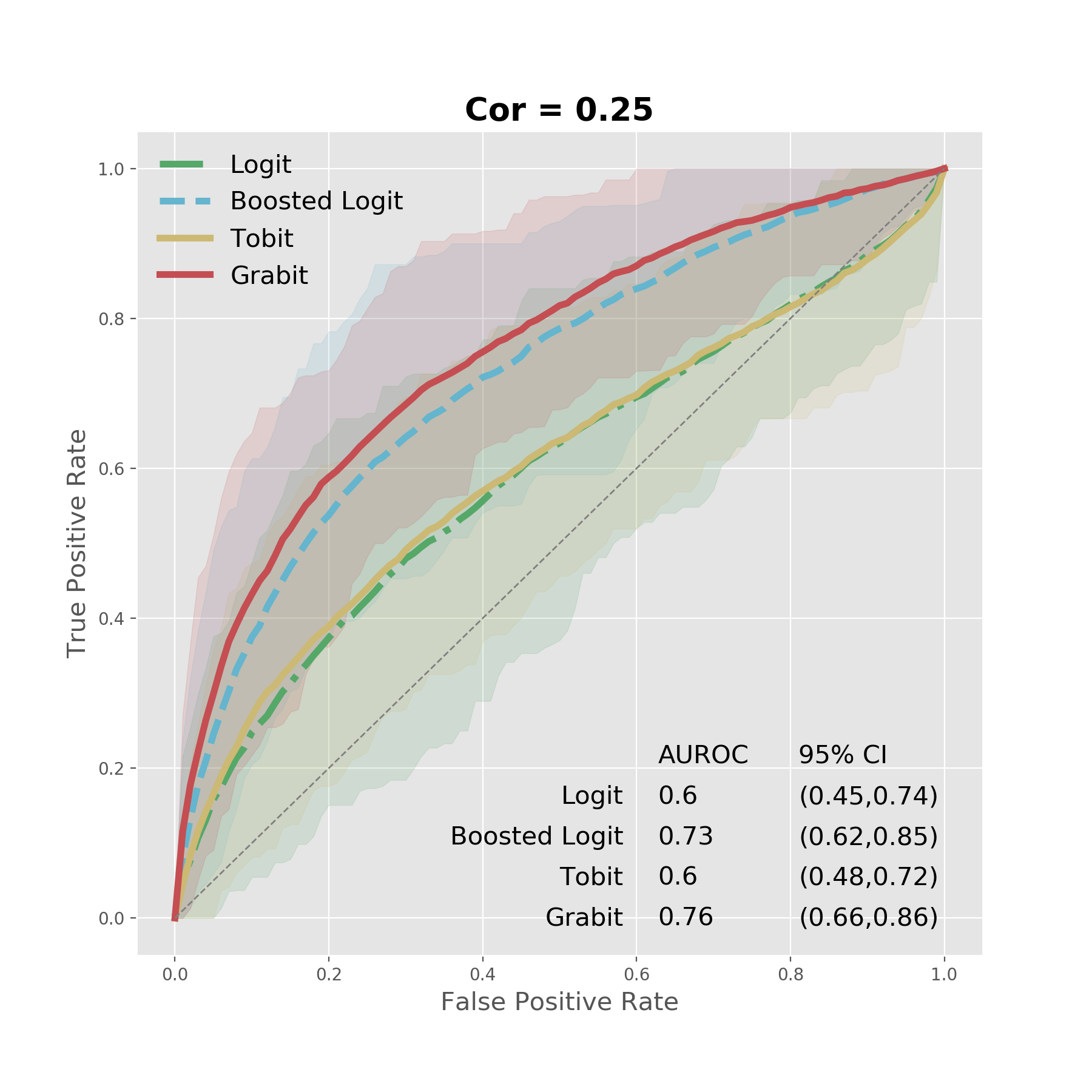}
	\includegraphics[width=0.49\textwidth,trim={1.25cm 1.5cm 1.25cm 1.5cm},clip=true]{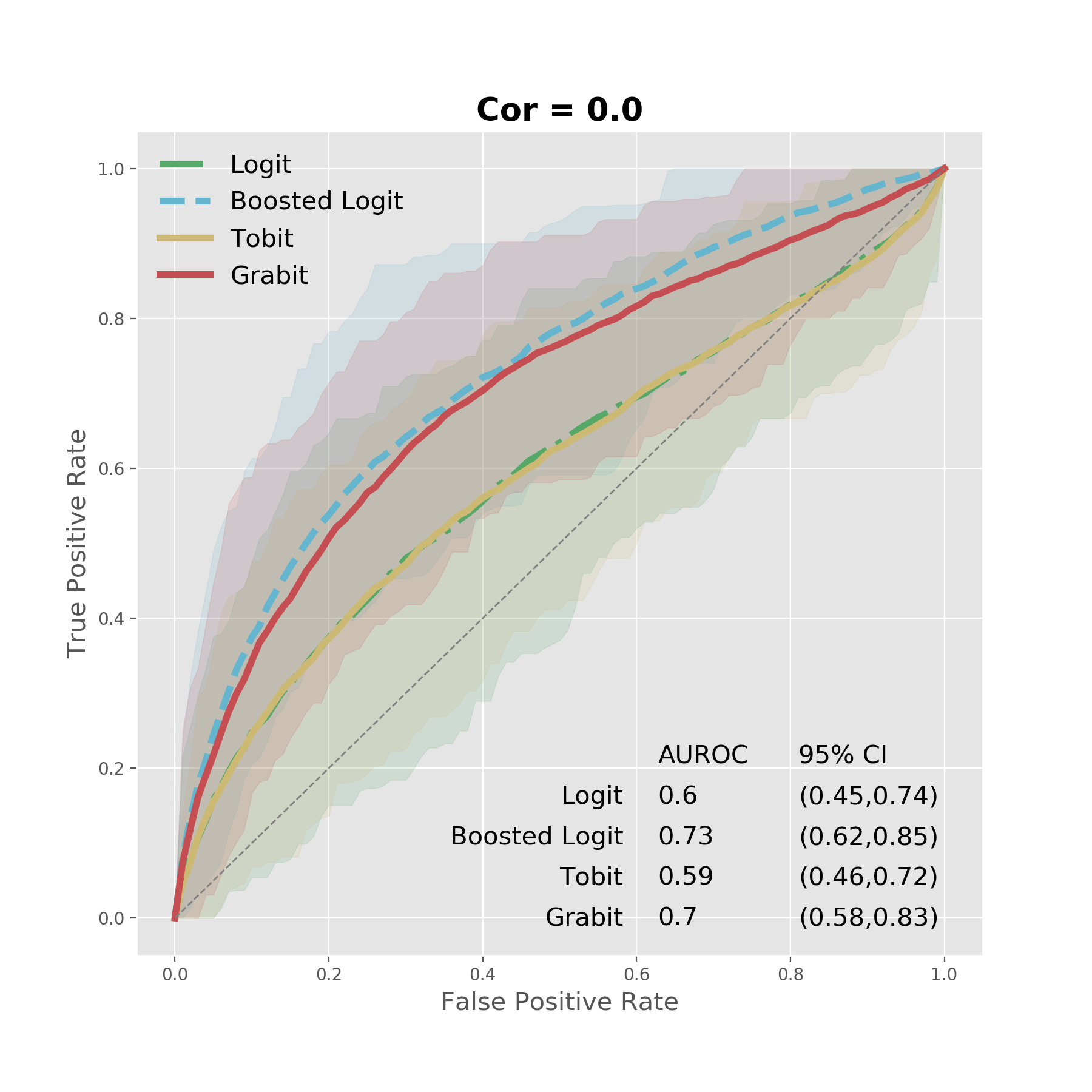}
	\caption{Comparison of models using receiver operating characteristic (ROC) and area under ROC (AUROC) for different levels of  correlation between the auxiliary data and the latent decision function when simulating from the model in \eqref{classsim} and \eqref{decfct}.}
	\label{fig:roc_sim} 
\end{figure}

In Figure \ref{fig:roc_sim}, we show the results for the four different correlation levels $0.75, 0.5, 0.25$, and $0$. We use the receiver operating characteristic (ROC) and the area under ROC (AUROC) to compare the different approaches. The AUROC values shown in the plots are sample means over the 100 simulations runs, and the 95\% confidence intervals for the AUROC are obtained by calculating 2.5\% and 97.5\% quantiles. ROC curves and 95\% confidence bands (shaded areas) are point-wise means as well as point-wise 2.5\% and 97.5\% quantiles of the 100 simulated ROC curves. Before calculating means and quantiles, the sample ROC curves are first linearly interpolated over an equally spaced grid with 100 grid points between 0 and 1.

The plots show that if the correlation between the auxiliary variable and the latent decision function is larger than zero, the Grabit model outperforms the other approaches. This means that the Grabit model can use the information in the auxiliary data in order to obtain higher predictive performance. Because of the nonlinearities and interactions in the function $F$ in \eqref{decfct}, the performances of both the Logit and the Tobit model are considerably worse. In the case where the correlation is zero, the predictive accuracy of the Grabit model is essentially equivalent to the one of the best performing competitor, i.e., the boosted Logit model. In this case where the auxiliary data contains no information, we lose no significant predictive accuracy when using the Grabit model. 


\subsection{Sample size}\label{sampsize}
\begin{figure}[ht!]
	\centering
	\includegraphics[width=0.49\textwidth,trim={1.25cm 1.5cm 1.25cm 1.5cm},clip=true]{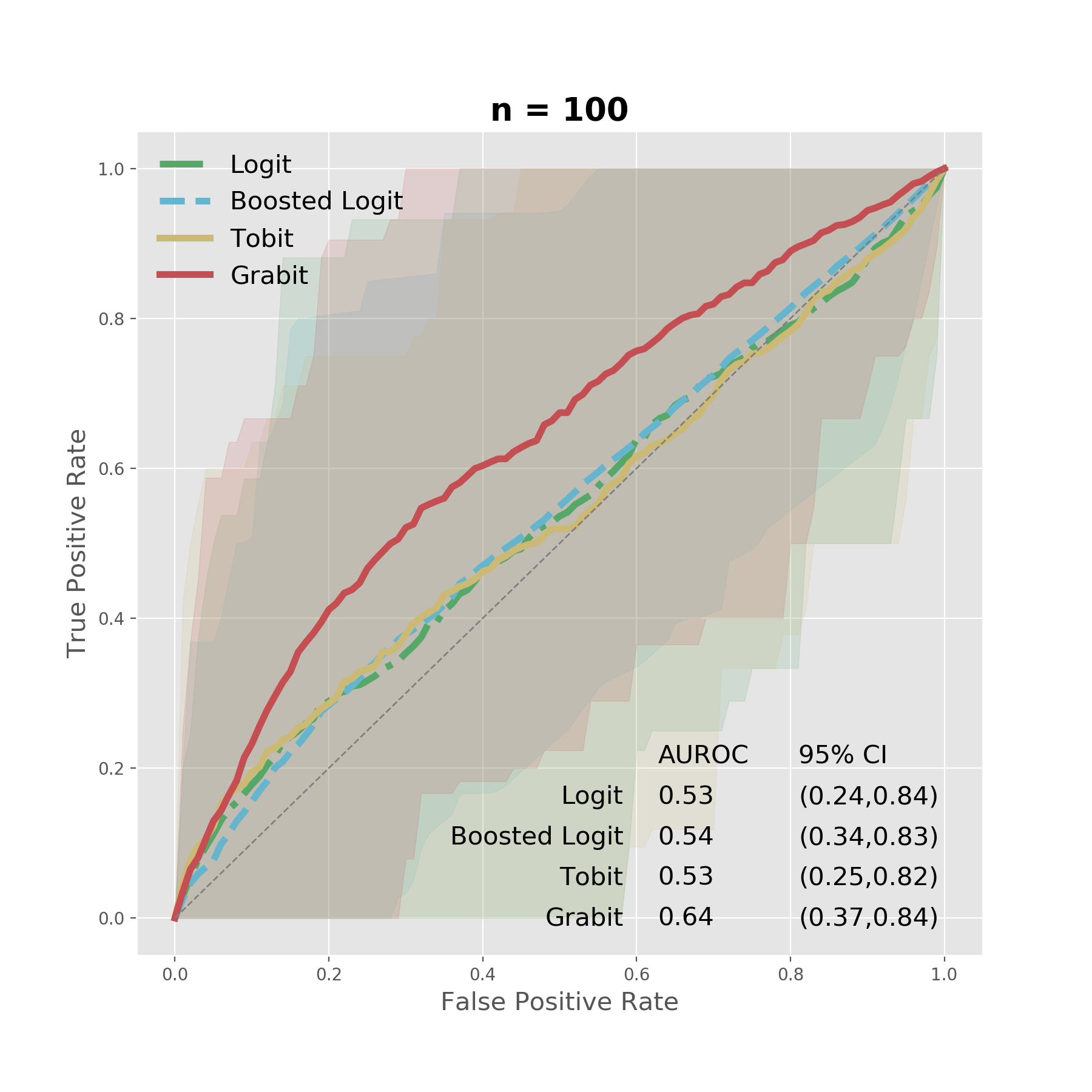}
	\includegraphics[width=0.49\textwidth,trim={1.25cm 1.5cm 1.25cm 1.5cm},clip=true]{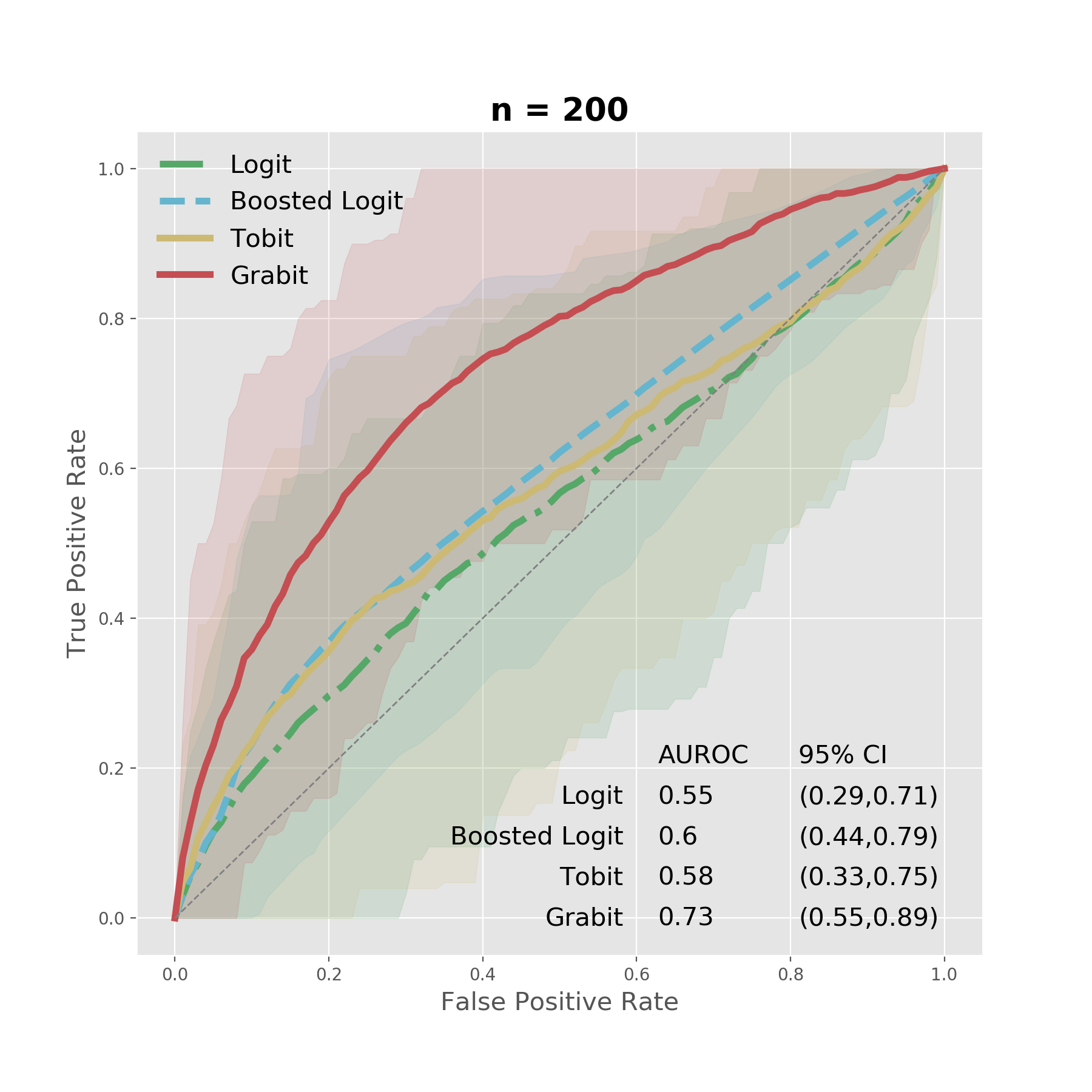}
	\includegraphics[width=0.49\textwidth,trim={1.25cm 1.5cm 1.25cm 1.5cm},clip=true]{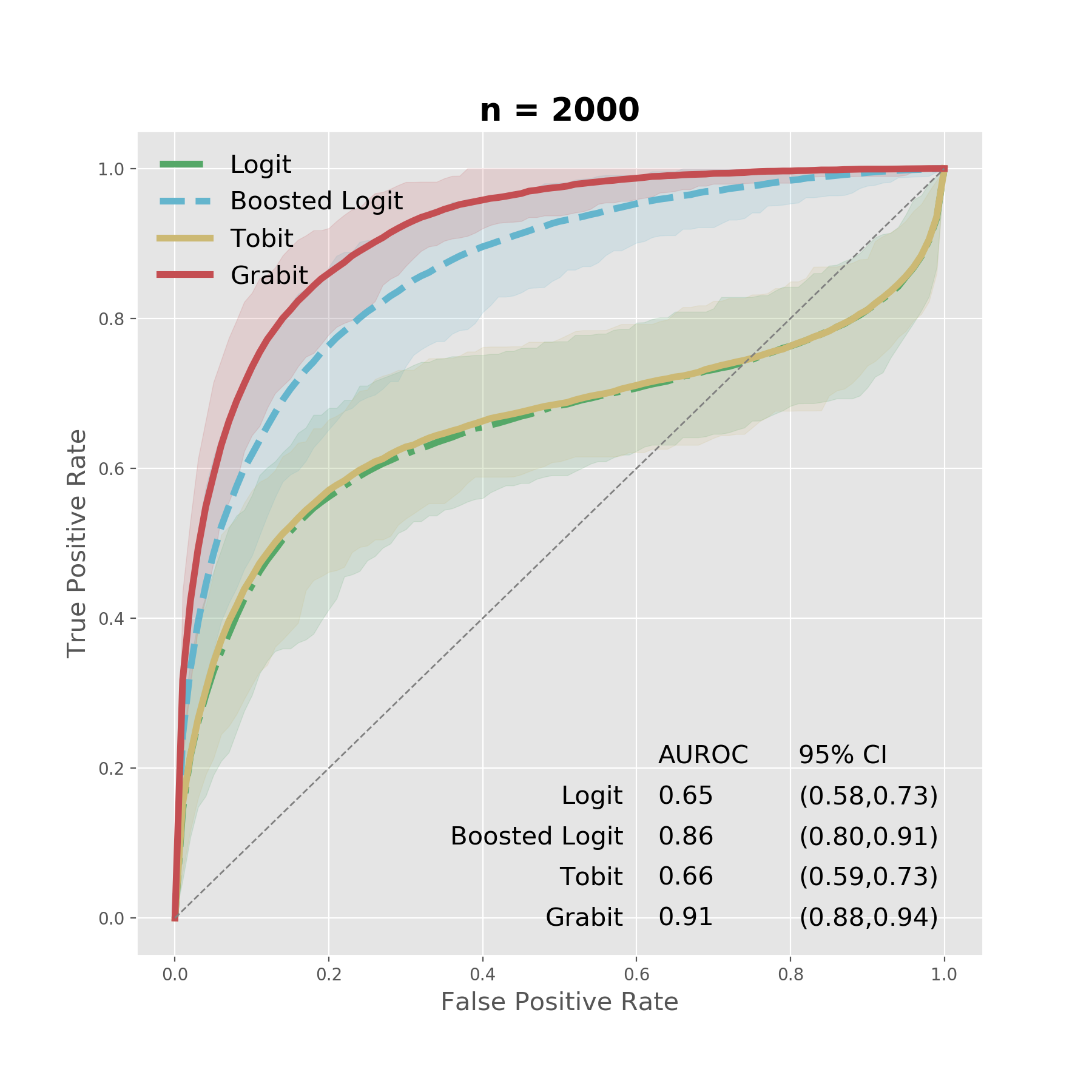}
	\includegraphics[width=0.49\textwidth,trim={1.25cm 1.5cm 1.25cm 1.5cm},clip=true]{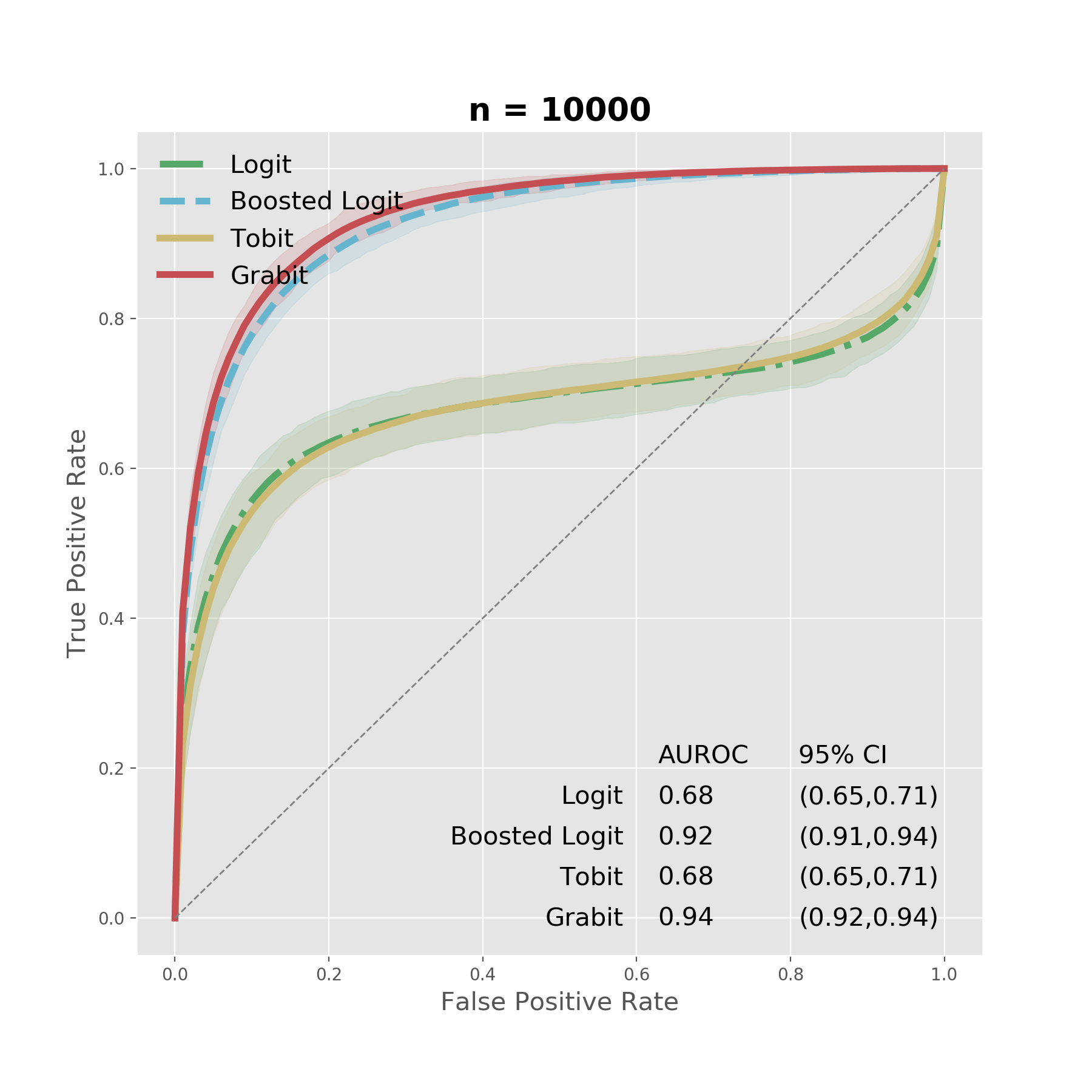}
	\caption{Results from a simulation study for investigating the impact of sample size on the performance of the Grabit model and other approaches.}
	\label{fig:roc_sim_n} 
\end{figure}

In this section, we investigate how the sample size impacts the performance of the Grabit model and the other classifiers. We use the same simulation setting as in Section \ref{simcor} with a correlation of $0.5$ between the auxiliary variable and the latent decision function. In addition to the sample size of $n=500$ from above, we consider the cases $n=100, 200, 2000$, and $10000$. The results are presented in Figure \ref{fig:roc_sim_n}. The results show that for smaller sample sizes, the performance gain of the Grabit model with respect to the other models is larger compared to larger data sizes. For the largest sample size, the performance of the boosted Logit model is almost as good as the one of the Grabit model. Again, due to the nonlinearity and interactions, both logistic regression and the Tobit model perform worse. As expected, the smaller the sample size, the wider the confidence bands for all methods.

\subsection{Class imbalance ratio}\label{classimbal}
We also explore the relation between the class imbalance ratio and the performance of the Grabit model and the other approaches. Previously, we have assumed that the minority class occurs in approximately $5$\% of all cases. Here, we additionally consider the following proportions of the minority class: $1$\%, $2$\%, $10$\%, and $20$\%.\footnote{This corresponds to using the following thresholds $y_u$: $3.89,3.44,2.38$, and $1.89$ in Equation \eqref{classsim}.} Apart from this, we use the same simulation setting as in Section \ref{simcor} with a correlation of $0.5$ between the auxiliary variable and the latent decision function.

\begin{figure}[ht!]
	\centering
	\includegraphics[width=0.49\textwidth,trim={1.25cm 1.5cm 1.25cm 1.5cm},clip=true]{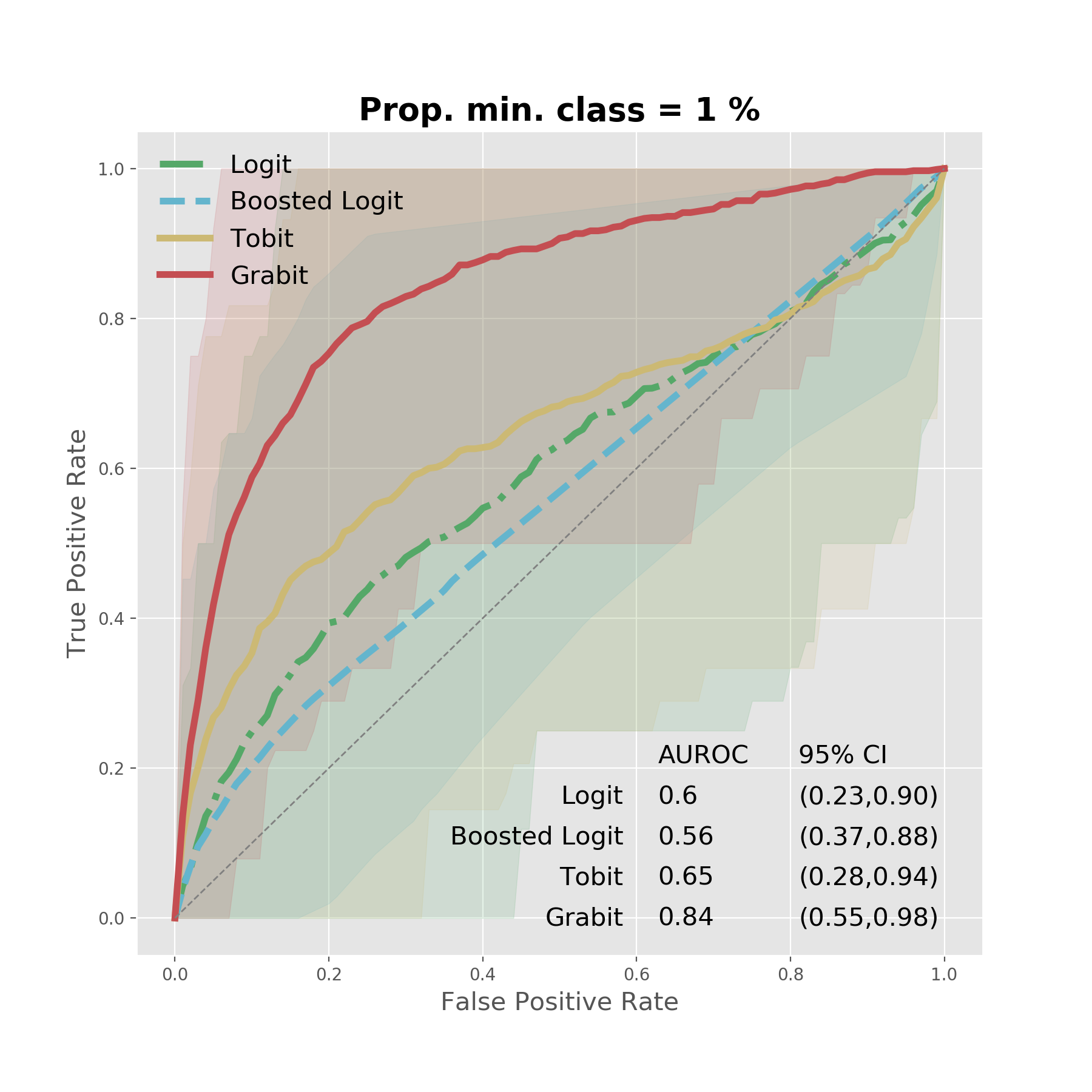}
	\includegraphics[width=0.49\textwidth,trim={1.25cm 1.5cm 1.25cm 1.5cm},clip=true]{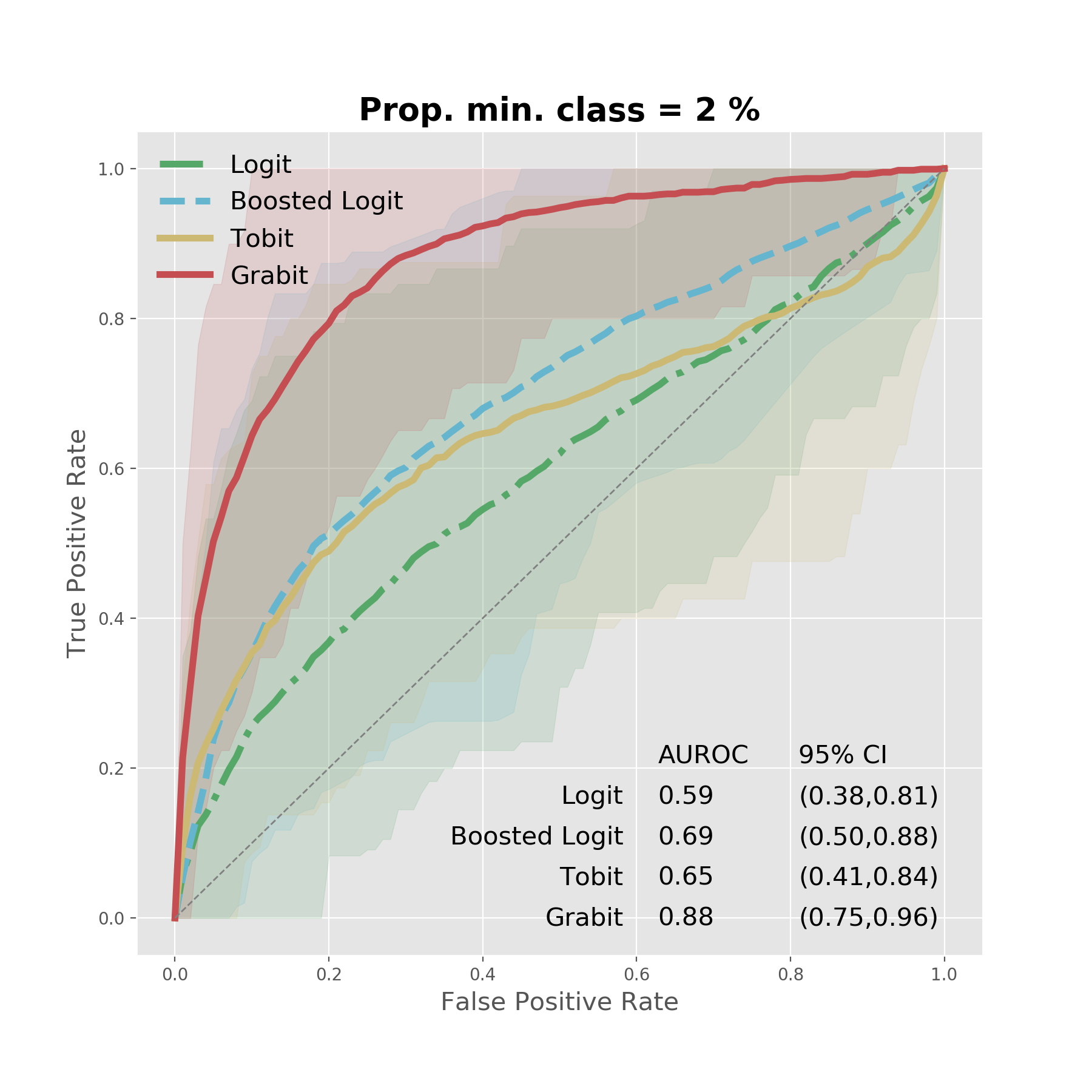}
	\includegraphics[width=0.49\textwidth,trim={1.25cm 1.5cm 1.25cm 1.5cm},clip=true]{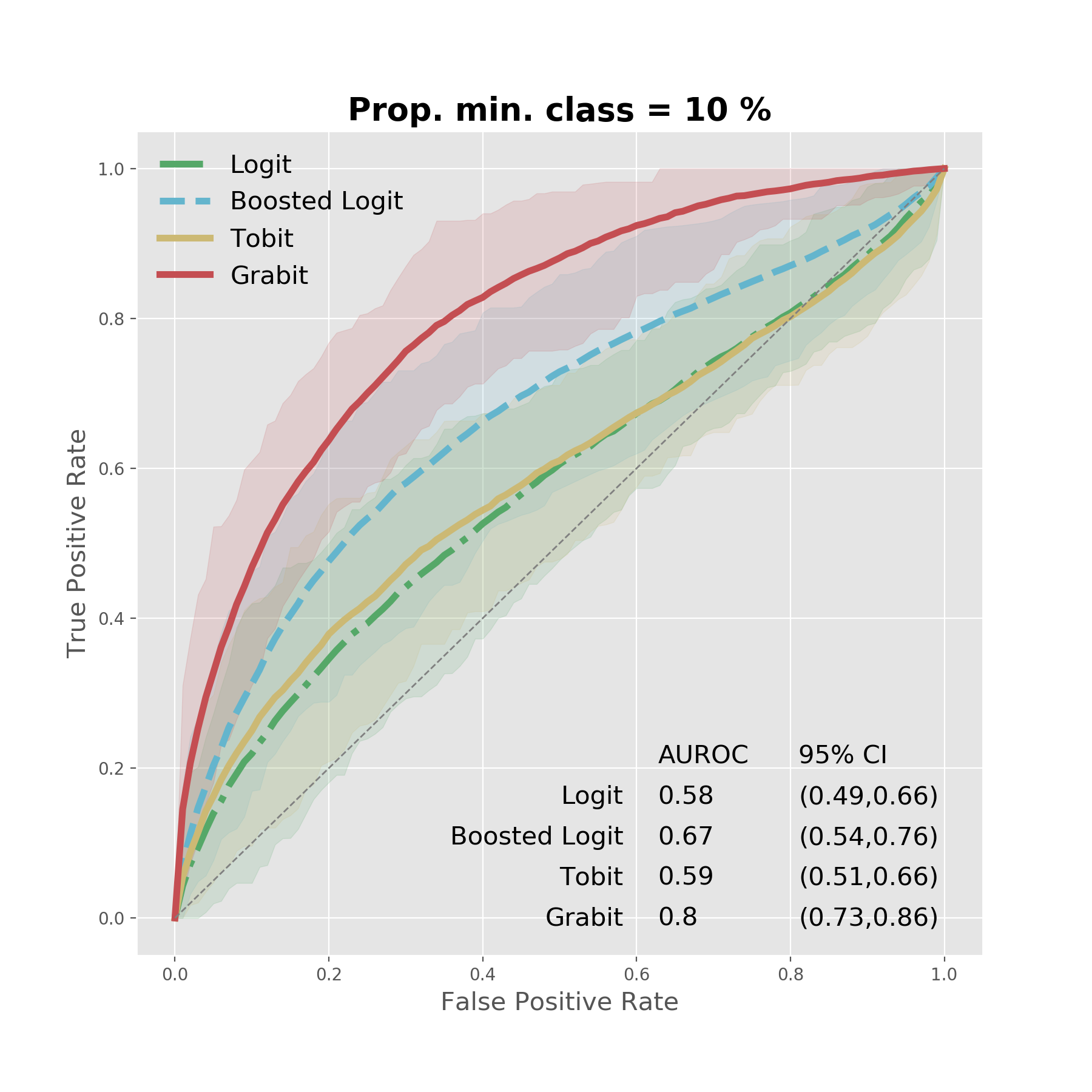}
	\includegraphics[width=0.49\textwidth,trim={1.25cm 1.5cm 1.25cm 1.5cm},clip=true]{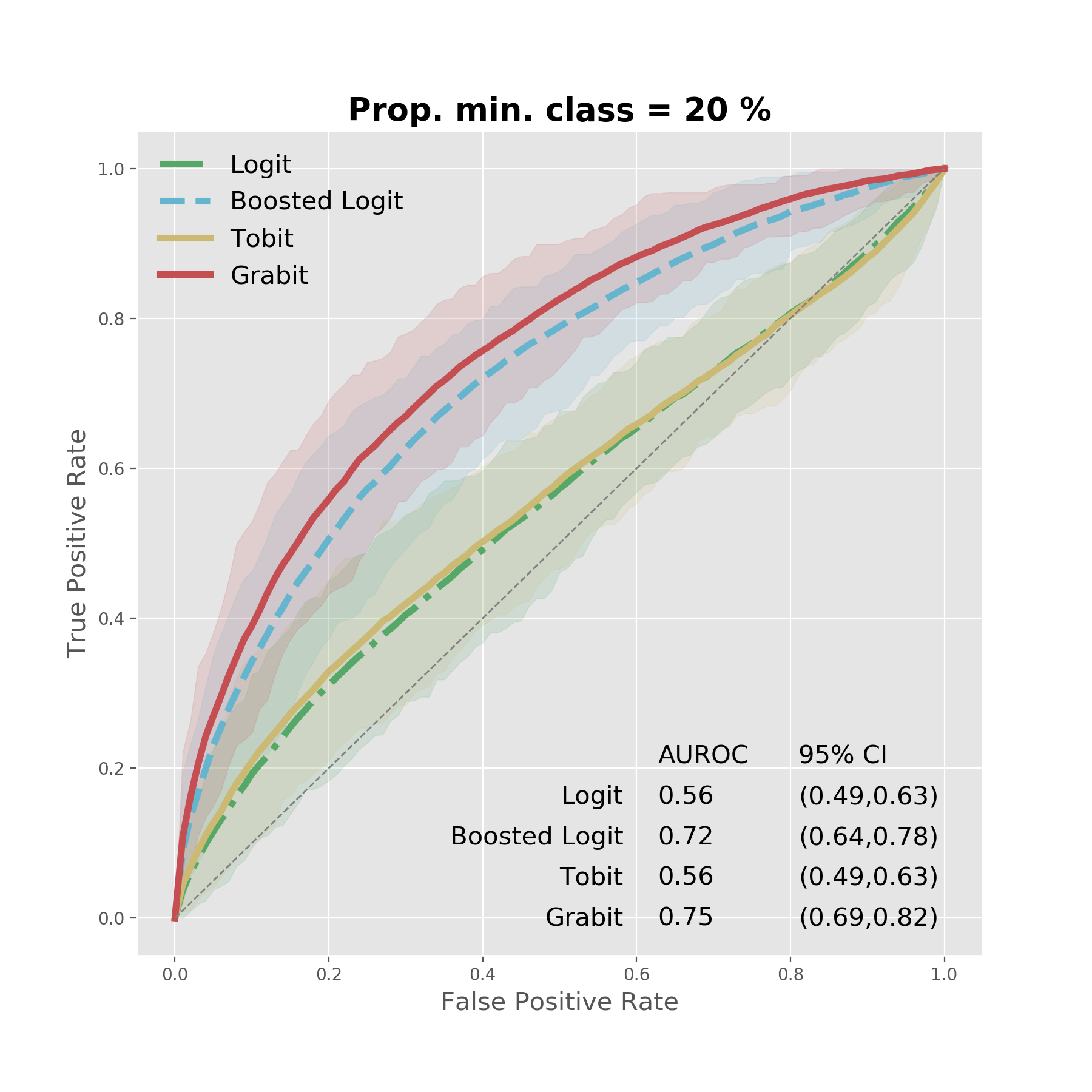}
	\caption{Results from a simulation study for investigating the impact of the class imbalance ratio on the performance of the Grabit model and other approaches.}
	\label{fig:roc_sim_DR} 
\end{figure}

The results are shown in Figure \ref{fig:roc_sim_DR}. We observe that the increase in predictive accuracy of the Grabit model compared to the boosted Logit model is larger the larger the class imbalance ratio or the lower the proportion of the minority class. As expected, the lower the fraction of the minority class, the wider the confidence bands due to the smaller number of minority class observations in both the training and test data. Note that we also observe that the performance of the Grabit model is slightly better for, e.g., the case of $2$\% minority samples compared to $20$\% minority samples. This is because it can be advantageous to directly observe a noisy version of the latent decision function compared to observing only a random binary variable with mean given by the decision function.

\subsection{Other decision functions}\label{decfunc}
In the following, we consider two other choices for the latent decision function $F$: a linear function and a highly nonlinear function in combination with a larger sample size. For the linear function, we replace \eqref{decfct} with
\begin{equation}\label{linfct}
F(X)=0.25\sum_{k=1}^{50} X_{k},~~X_k \overset{\text{ iid }}{\sim} \text{Unif}(-1,1).
\end{equation} 

As nonlinear function, we use the following choice
\begin{equation}\label{nonlinfct}
F(X)=2\cos\left(4\pi\sqrt{\sum_{k=1}^{20}X_k^2} \right),~~X_k \overset{\text{ iid }}{\sim} \text{Unif}(-1,1).
\end{equation} 
The reason why we choose such a highly nonlinear function is not to mimic typical relationships expected in credit risk applications, but rather to investigate whether the Grabit model can also provide increased predictive accuracy in cases where the sample size is larger and the decision function is of complex nature. 

In both cases, $y_u$ is chosen such that approximately 5\% of all cases are in the minority class, $\sigma_a$ is chosen such that the correlation between the auxiliary variable and the latent decision function $F$ is approximately $0.5$, and $\mu_a$ such that all simulated auxiliary data is below $y_u$. For the linear model, we simulate $n=500$ data points and for the nonlinear function, we simulate $n=10000$ samples for both training and test data in each of the $100$ simulation iterations.

\begin{figure}[ht!]
	\centering
	\includegraphics[width=0.49\textwidth,trim={1.25cm 1.5cm 1.25cm 1.5cm},clip=true]{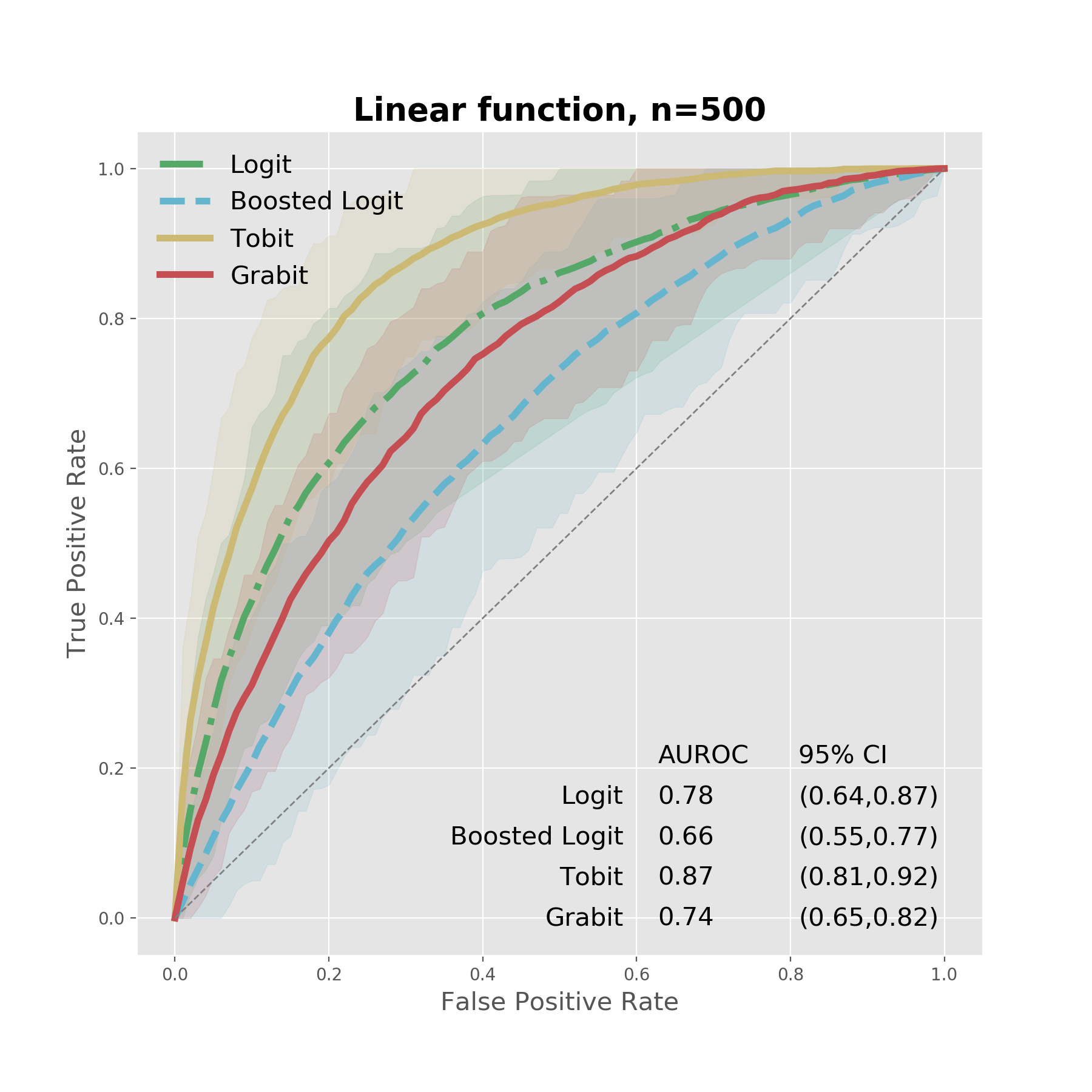}
	\includegraphics[width=0.49\textwidth,trim={1.25cm 1.5cm 1.25cm 1.5cm},clip=true]{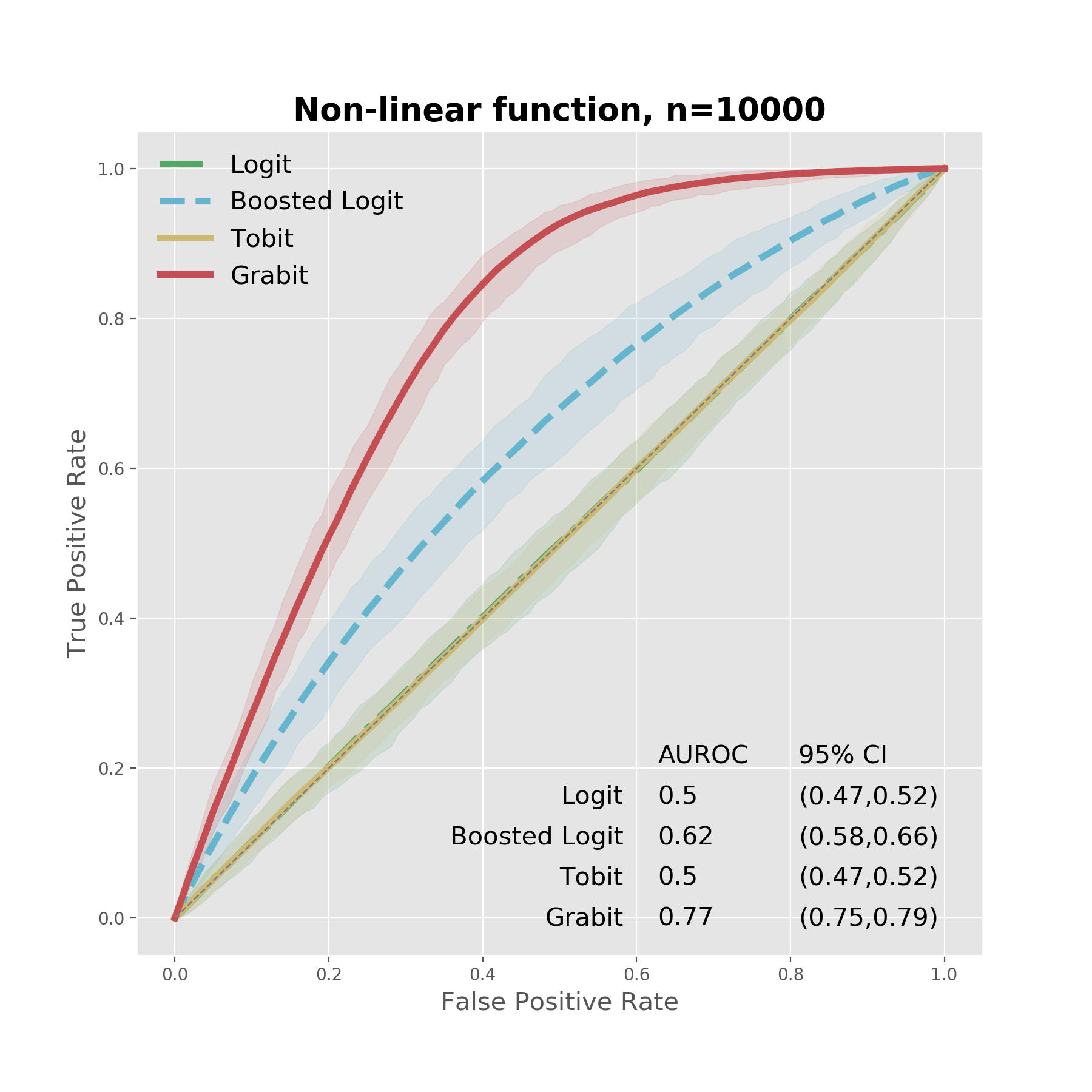}
	\caption{Results from a simulation study for investigating the impact of the structure of the latent decision function and the sample size on the performance of the Grabit model and other approaches.}
	\label{fig:roc_sim_other} 
\end{figure}

The results are shown in Figure \ref{fig:roc_sim_other}. As expected, the figure shows that in the linear case, the Tobit model performs best. In particular, the Tobit model outperforms the tree based Grabit model. It is not surprising that linear models perform best in situations where the true decision function is linear. Similarly, the tree-boosted Logit model performs worse than the linear Logit model. Nonetheless, it is beneficial to use the auxiliary data as both the Tobit and the Grabit model outperform their binary classifier counterparts, i.e., the Logit and the boosted Logit model.

For the nonlinear function in \eqref{nonlinfct} with $n=10000$ simulated data points, we observe that the Grabit model clearly outperforms the other three approaches. For the relatively simple decision function in Equation \eqref{decfct}, the results reported in Section \ref{sampsize} show that with a sample size of $10000$, the boosted Logit model is almost as accurate as the Grabit model. This example thus shows that the Grabit model can also provide increased predictive accuracy for moderately- to large-sized datasets when the decision function is sufficiently complex, e.g., having strong nonlinearities, interactions, or many predictor variables.

In summary, we find the following results in this simulation study. First, the larger the correlation between the auxiliary data and the latent decision function, the larger is the performance gain of the Grabit model compared to other classification methods, which only use binary data and neglect the auxiliary data. In the case of zero correlation, the Grabit model can perform as good as it second best competitor, the boosted Logit model. Further, the larger the class imbalance or the smaller the sample size, the larger the performance gain of the Grabit model. Finally, the Grabit model outperforms other models also in cases of larger datasets if the decision function is complex such as, e.g., showing strong nonlinearities or interactions among predictors. 


\section{Application to SME default prediction}\label{defaulpred}
In this section, we apply the Grabit model to default prediction of loans made to small and medium-sized enterprises (SME) in Switzerland. The data is provided by Advanon, a Swiss start-up company, which operates a platform on which SMEs can obtain short-term loans by pre-financing their invoices. The goal of this application is to predict whether a loan will be repaid or not when an SME requests a new loan.

On Advanon's platform, a loan can be repaid with some delay without immediately being classified as a default. This means that in addition to the information whether a company defaulted on a loan or not, we know for each loan whether it was repaid in due time or, if not, the number of days of delay by which it was repaid. The maximum number of days in arrear is 60. In case a loan is overdue more than 60 days, the loan is automatically classified as a default event by Advanon. 

We use the Grabit model for jointly modeling the auxiliary delay days and the binary default events. The observed variable $Y$ is a censored version of a latent variable $Y^*$, where the latter can be interpreted as a default potential or an inverse credit score. Both lower and upper censoring occur at $y_l=0$ and $y_u=60$, respectively. If the latent variable exceeds the upper threshold, $Y^*\geq 60$, a default occurs. This means that all default events correspond to $Y=60$. If the latent variable is below $60$, $Y^*<60$, no default occurs, and the observed $Y$ equals the number of days of delay. Lower censoring at $y_l=0$ is introduced in order to account for the large number of loans that were repaid without any delay.

\subsection{Data}
\begin{figure}[ht!]
	\centering
	\includegraphics[width=0.8\textwidth]{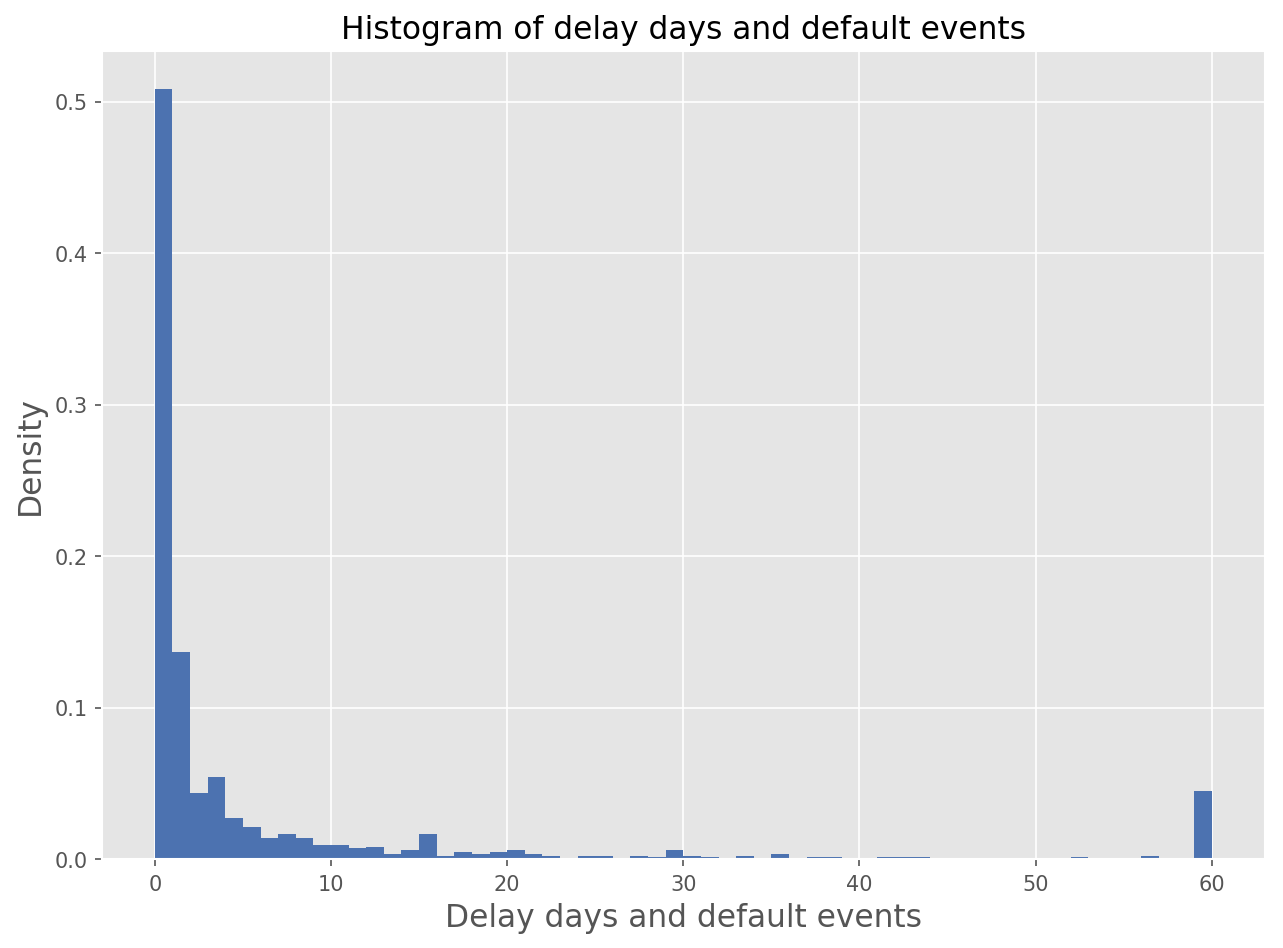}
	\caption{Histogram of delay days and default events. The bar at 60 does not correspond to delay days but represents the fraction of default events. The remaining part of the histogram below 60 represents the number of delay days.} 
	\label{fig:hist} 
\end{figure}

The data consist of 850 loans made to 141 different Swiss SMEs between 2016 and 2017. In total, 36 loans were not repaid due to defaults of 14 different SMEs. Note that a company can request more than one loan at the same time, and the lifespan of different loans for the same company can be overlapping.  In Figure \ref{fig:hist}, we illustrate the default events and the number of delay days by which loans were repaid. The point mass at 60 represents the fraction of default events and the remaining part of the histogram below 60 represents delay days.

For each loan, there are approximately 50 different predictor variables. These covariates include financial ratios calculated from balance sheets and income statements, SME characteristics such as variables that reflect the repayment history of the SME on the platform or the age of the company, loan characteristics such as the loan amount or time until maturity, ratings for the SME from social media platforms, data from several external rating agencies, and data about online user behavior such as log-in and click data. We transform variables that are highly skewed such as loan amounts or several balance sheet summaries using a logarithmic transformation. We do this in order to mitigate the influence of single data points when applying a logistic regression model for comparison below. For tree-based models such as the Grabit model, this is not needed since such models are invariant under monotone transformation. 


For confidentiality reasons, we cannot fully disclose all predictor variables, and the data used here consists of a random subsample of all loans made on Advanon's platform to Swiss SMEs. The subsample contains all default events but only a random selection of all non-defaulted loans. This is done in order to not disclose the actual default rate on Advanon's platform, which is different from the one in the random subsample used here. However, the results shown in the following change only marginally when using the full dataset.

\subsection{Model evaluation and comparison}\label{eval_appl}
We compare the performance of the Grabit model with several alternatives: logistic regression ("Logit"), a classification tree, a random forest, tree-boosted logistic regression ("boosted Logit"), a neural network, the Tobit model, and a tree-boosted multinomial logistic regression model ("boosted multiclass Logit"). Apart from the Tobit model and the boosted multiclass Logit model, the methods are classification techniques that only use the binary information whether a loan was repaid or not. Similarly as the Grabit model, the Tobit model also uses the number of delay days as auxiliary data but assumes a linear decision function. The boosted multiclass Logit model uses the auxiliary data as follows. We create four states: "no delay", a delay between $1$ and $30$ days, a delay between $31$ and $60$ days, and default. We then use a boosted multinomial softmax classifier as described in \citet{friedman2001greedy} to model these four discrete states. Concerning the classification tree and the random forest algorithm, see Appendix \ref{tree_rf} for a short description. Except for the Grabit model and the Tobit model, the Python package \texttt{scikit-learn} \citep{scikit-learn} is used for fitting the models.  Tuning parameters are chosen using the temporal cross-validation scheme described in the following with the area under the receiver operating characteristic (AUROC) as a measure of fit. See Appendix \ref{tune_pars} for more details.

For evaluating and comparing the different models, we use temporal cross-validation. For each loan, we make a prediction and compare the prediction with the actual outcome. In doing so, the models are estimated based on data of past loans only, i.e., loans that are either repaid or have been classified as default events at the time when making the prediction. To avoid a temporal censoring issue, we only use data for loans for which the repayment date is at least 61 days due at the time of estimation. In addition, we require at least 100 past data points in order to train a model and to make a prediction. In total, there are 610 loans, out of which 28 are default cases, for which we can make this temporal out-of-sample evaluation. For a small fraction of the data, there are missing values in some of the predictor variables. These are simply interpolated by using the median of all past data available at that point in time. 

\begin{figure}[h!]
	\centering
	\includegraphics[width=0.9\textwidth]{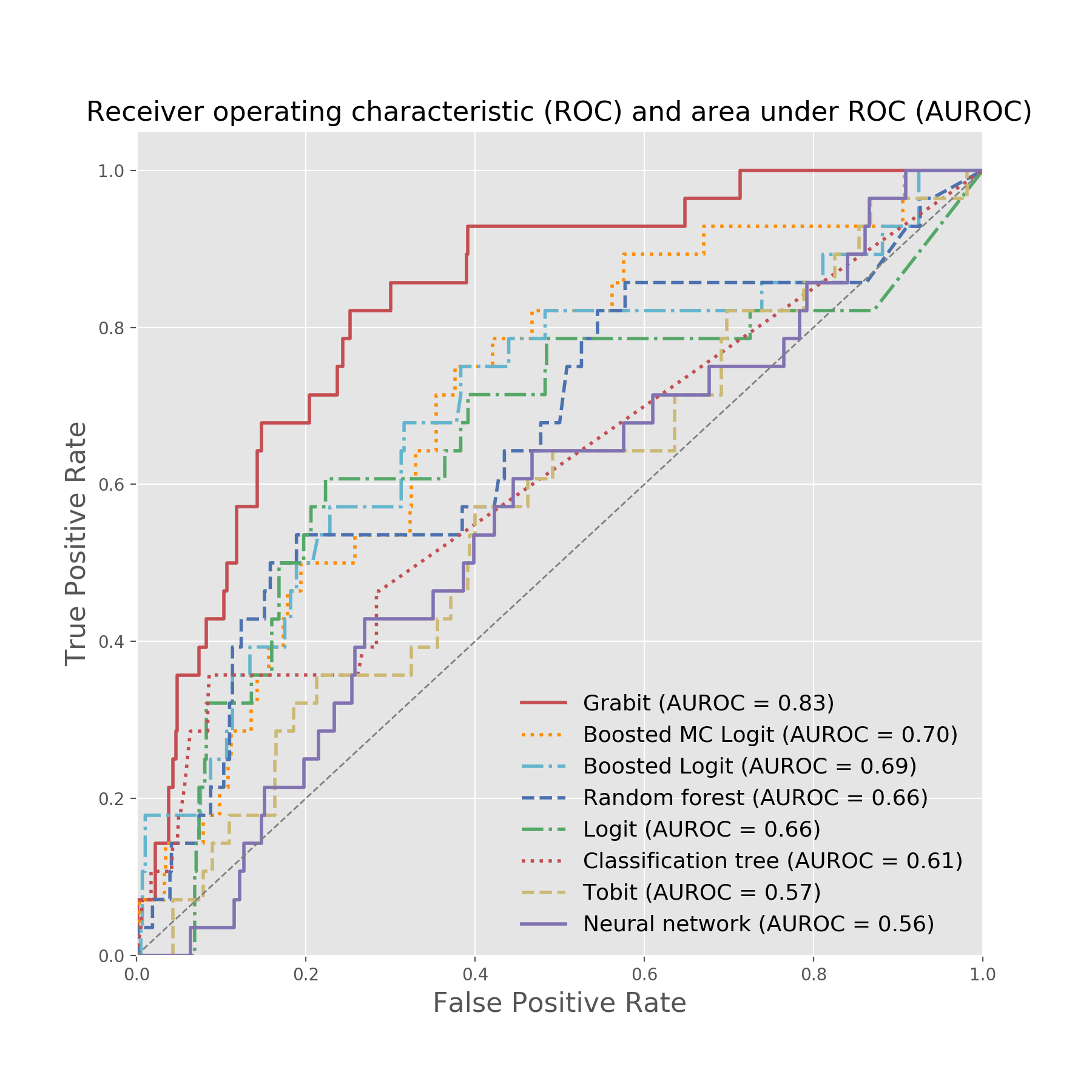}
	\caption{Comparison of different models using receiver operating characteristic (ROC) and area under ROC (AUROC) using temporal cross-validation.}
	\label{fig:roc} 
\end{figure}

We evaluate the predictions and compare the different models using the receiver operating characteristic (ROC) curve and the area under ROC (AUROC). The ROC curve is obtained by plotting the true positive rate versus the false positive rate for varying thresholds. The AUROC is the area under this curve. Figure \ref{fig:roc} shows our results. As the figure shows, the Grabit model clearly outperforms all other approaches. For most of the thresholds, the ROC curve of the Grabit model is above the ROC curves of all other models. Further, the AUROC is considerably larger compared to all other alternative models considered. In particular, the AUROC of the Grabit model is significantly higher at the 5\% level compared to all other models when using the DeLong test \citep{delong1988comparing}. The corresponding p-values are reported in Table \ref{tab:DeLong}.  Figure \ref{fig:roc} also shows that the Tobit model has low predictive accuracy. Similarly as in the simulation study in Section \ref{simstudy}, this is likely due to the presence of nonlinearity and interactions. 

\begin{table}[h!]
	\begin{center}
		\caption{Comparison of AUROC of the Grabit model with alternative approaches using the DeLong test.}
		\label{tab:DeLong}
		\begin{tabular}{lc}
			\hline    \hline
			Model & p-Value\\
			\hline
			Boosted Logit  & 0.0371\\
			Boosted multiclass Logit &0.0379 \\
			Classification tree & 0.000173\\
			Logit & 0.002\\
			Neural network& 2.83e-06\\
			Random forest  & 0.00128\\
			Tobit  & 3.27e-07\\
			\hline
		\end{tabular}
	\end{center}
\end{table}

Due to the small sample size, we use the same data for choosing the tuning parameters and for comparing the different models, i.e., we do not distinguish between a validation and test data. This comes at the potential risk that the results of methods for which tuning parameters are chosen in this way are too optimistic. In order to investigate whether this is an issue, we also consider other non-optimal choices of tuning parameters. In particular, in Figure \ref{fig:roc_non_opt} in Appendix \ref{tune_pars}, we report the results when using the second best choice of tuning parameters for the boosting methods. The performance of the Grabit model is essentially the same whereas some of the other methods perform marginally worse. We obtain similar results when using other non-optimal, but reasonable, tuning parameters (results not tabulated). These results show that our findings are robust to different choices of tuning parameters. In addition, concerning the use of parameters optimally tuned on the test data for boosting methods, \citet{buhlmann2003boosting} state that "the effect of using the optimal number of boosting iterations instead of an estimated [e.g. by cross-validation] number is typically small".

An alternative approach for dealing with imbalanced binary data is over- or under-sampling as well as synthetic minority over-sampling technique (SMOTE) \citep{chawla2002smote}. In our case, these approaches do not improve the accuracy of the binary classifiers that we consider (results not tabulated). The likely reason for this is that the Grabit model can learn additional structure from the auxiliary data that is not present in the binary data. In contrast, over- and under-sampling or also SMOTE, cannot use the auxiliary data and, consequently, cannot learn this structure.

Further, note that we implicitly assume that multiple loans from the same borrower are independent conditional on the predictor variables. We do, however, account for borrower specific dependence by including, e.g., the repayment history of a company such as a repayment score, the number of loans repaid, the prior total amount repaid, and the maximum past delays in the covariates. Even if there is a certain amount of additional residual dependence, we believe that this is not a problem in our case for the following reasons. First, given that in linear regression models, coefficient estimates are unbiased also in the presence of correlation, it is likely that the learned mean function in the Grabit model is also unbiased and not impacted by correlation. Further, our goal in this application is univariate default prediction. If the goal is the prediction of aggregate default rates of, e.g., loan portfolios, over a longer time period this assumption might not be applicable. In this case, a simple solution for prediction is to assume implicit cross-default provisions and declare all loans as nonperforming once a borrower is predicted to default on one of its loans. Finally, the reason why we use loans and not borrowers as observation units is that, given the small sample size, it is not possible to aggregate to the borrower level at, e.g., yearly frequency. However, the distribution of the number of loans per defaulted borrower is not skewed, and we do not have the problem that a few borrowers with many loans might disproportionately influence the results. Nonetheless, we have also considered loan amount weighed ROCs and AUROCs and obtain similar performance gains with the Grabit model compared to the other approaches (results not tabulated).

\subsection{Model interpretation}
In general, interpretability is an important factor for the adoption of machine learning models in financial applications. Currently, the lack of interpretability often hampers the adoption of complex machine learning approaches. Linear models such as logistic regression have the advantage that they can be easily interpreted. The Grabit model, and boosted trees in general, are not as easily interpretable as linear models as such models consist of an ensemble of a large number of trees. However, compared to other nonlinear models such as neural networks or support vector machines, boosted trees can be relatively well interpreted. In this section, we briefly illustrate two tools for model interpretation, variable importance measures and partial dependence plots, as well as an approach for explaining predictions, which we denote as `local partial dependence plots'.



\subsubsection{Variable importance}
A variable importance measure quantifies the importance of single variables for the prediction of $Y$. For a single $J$-terminal node tree $T^{[m]}$, \citet{breiman1984classification} proposed the following measure of importance for variable $X_l$
$$\mathcal{I}_{X_l}\left(T^{[m]}\right)=\sum_{j=1}^{J-1}\widehat{\imath}^2_j\mathbbm{1}_{v_j}(X_j), $$
where the sum is over all nonterminal nodes $j$, $v_j$ is the splitting variable selected in node $j$, $\mathbbm{1}_{v_j}(X_j)$ is an indicator function that equals one if $X_j$ is the splitting variable in node $j$ and zero otherwise, and $\widehat{\imath}^2_j$ denotes the reduction in squared error due to split $j$.

\citet{friedman2001greedy} generalized this measure to an ensemble of $M$ trees obtained by boosting by simply taking the average of the measures of all single trees
$$ \mathcal{I}_{X_l}=\frac{1}{M}\sum_{m=1}^M\mathcal{I}_{X_l}\left(T^{[m]}\right).$$
We note that $\mathcal{I}_{X_l}\left(T^{[m]}\right)$ and $\mathcal{I}_{X_l}$ can be biased \citep{breiman1984classification} in the sense that a variable $X_l$ that is independent of $Y$ might still be selected for a split in a tree, and hence the variable importance measure $\mathcal{I}_{X_l}$ might not be zero. A bias correction can be obtained as presented in \citet{sandri2008bias}.

\begin{figure}[ht!]
	\centering
	\includegraphics[width=\textwidth]{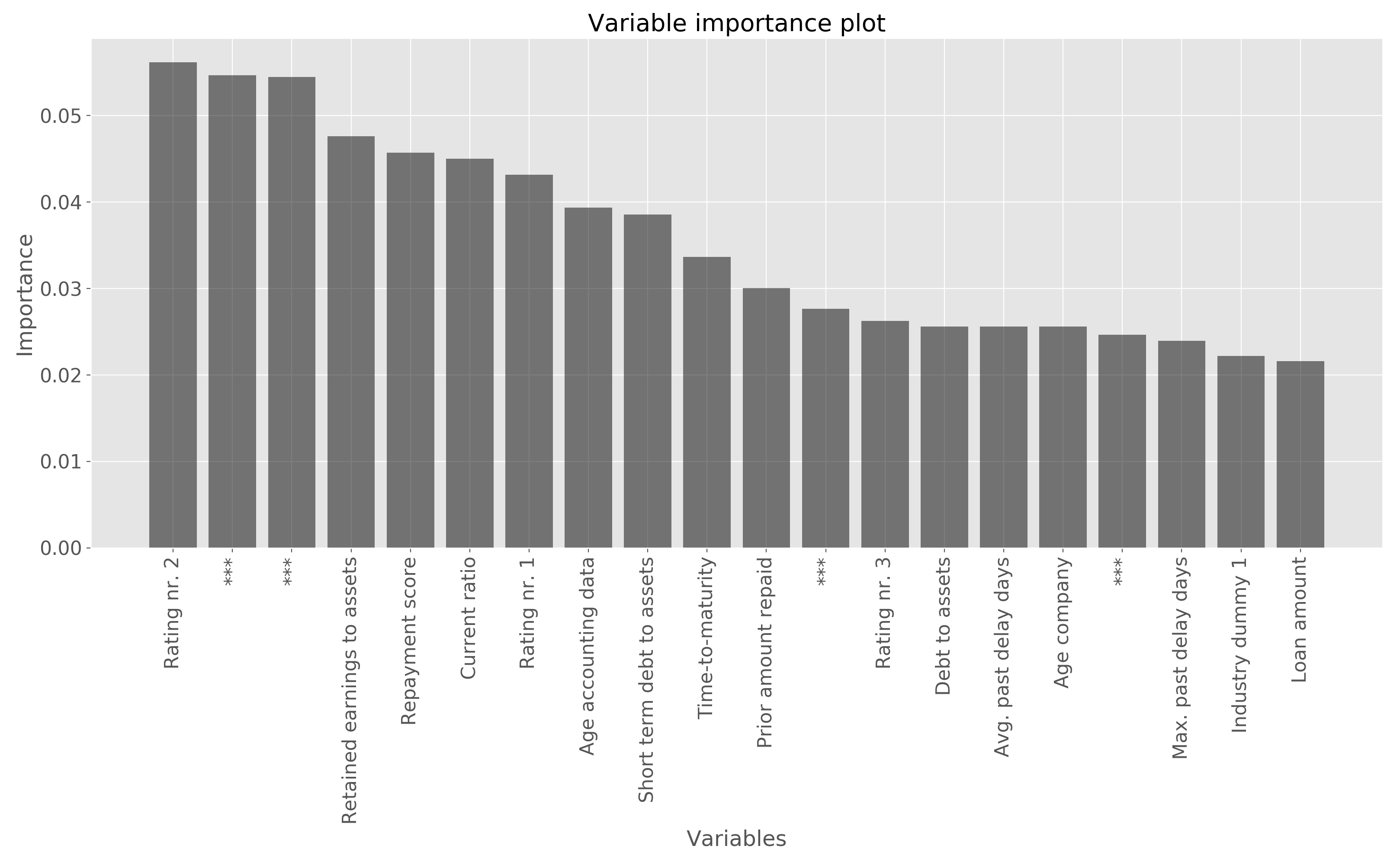}
	\caption{Variable importances for the 20 most important variables.}
	\label{fig:VarImp} 
\end{figure}
In Figure \ref{fig:VarImp}, we show variable importance measures for the 20 most important variables. Note that due to confidentiality, the names of a few the variables cannot be disclosed. Among the most important variables are ratings from external rating providers, various financial ratios, as well as measures of past repayment behavior such as a repayment score calculated by Advanon or the maximum number of days of delay by which loans were repaid in the past. 


\subsubsection{Partial dependence plots}\label{partdep}
Visualization is a powerful tool for interpreting models. For boosted trees, one can visualize main effects and second-order interactions using partial dependence plots \citep{friedman2001greedy}. For this, one partitions the predictor variables $X$ in two non-overlapping subsets $X_s$ and $X_{-s}$, where $X_{-s}$ is the complement of $X_s$. For main effects and second-order interactions, $X_s$ simply consists of one or two variables. Given a model $\widehat{F}(\cdot)$ and data $(y_i,x_i)$, an estimate for the average partial dependence of $\widehat{F}(\cdot)$ on $X_s$ can be calculated as
\begin{equation}\label{partdepplot}
\widehat{ \overline{F}}_s(X_s)=\frac{1}{n}\sum_{i=1}^n\widehat{F}(X_s,x_{-s,i}).
\end{equation}
A partial dependence plot is then obtained by plotting $\widehat{ \overline{F}}_s(X_s)$ versus $X_s$.

\begin{figure}[ht!]
	\centering
	\includegraphics[width=0.9\textwidth]{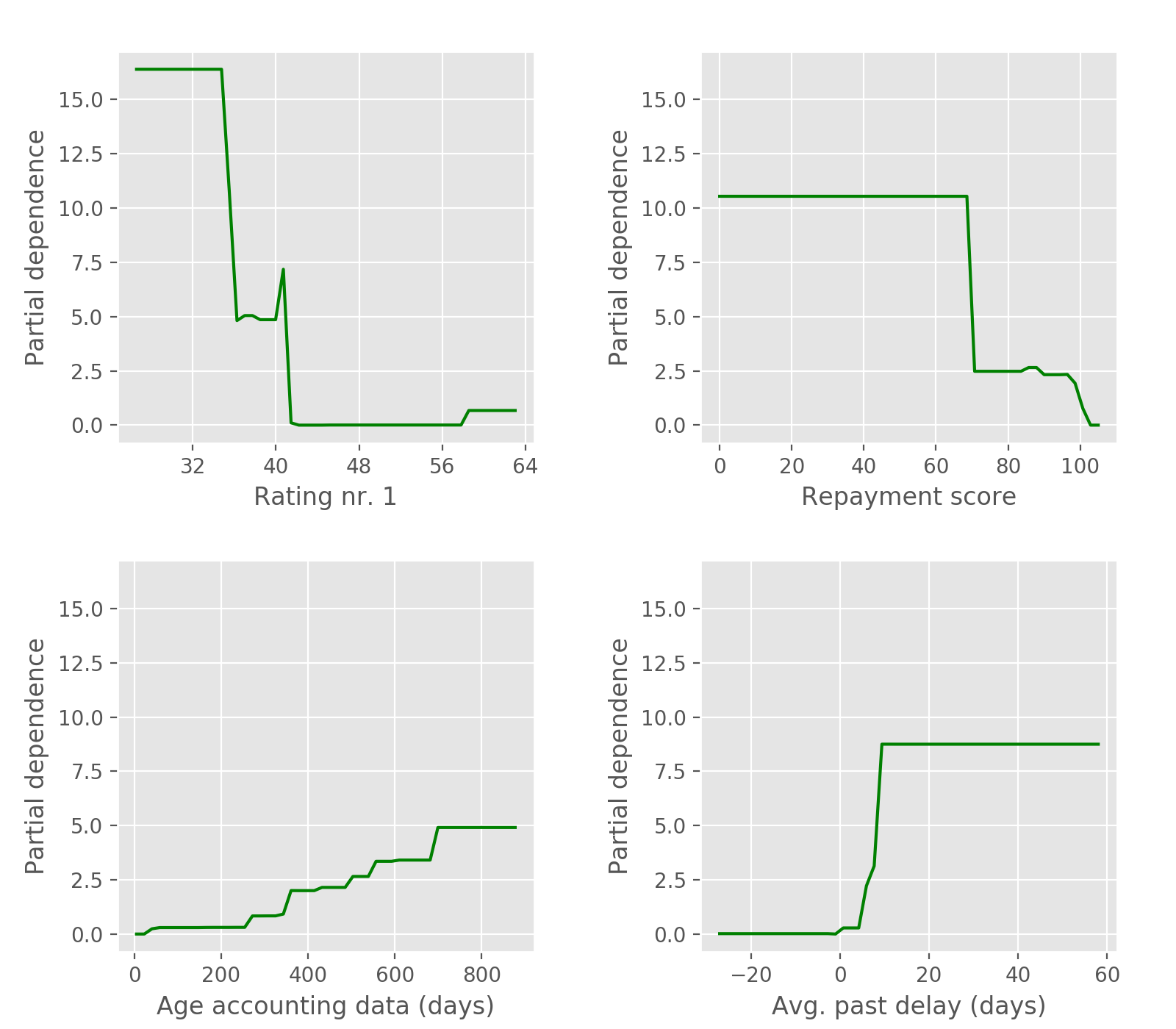}
	\caption{Partial dependence plots illustrating main effects for four variables with high variable importance.}
	\label{fig:imp1} 
\end{figure}

\begin{figure}[ht!]
	\centering
	\includegraphics[width=0.49\textwidth]{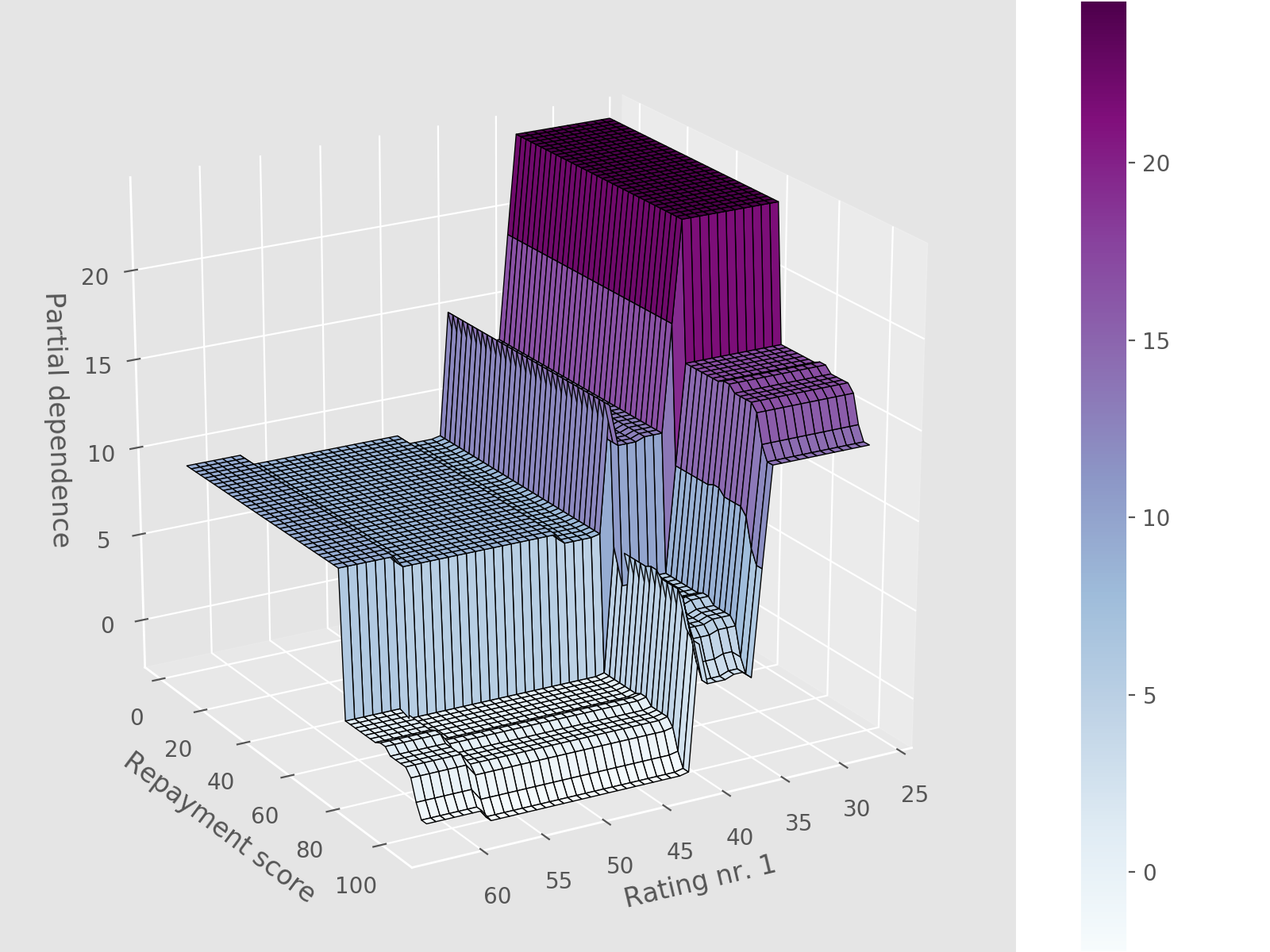}
	\includegraphics[width=0.49\textwidth]{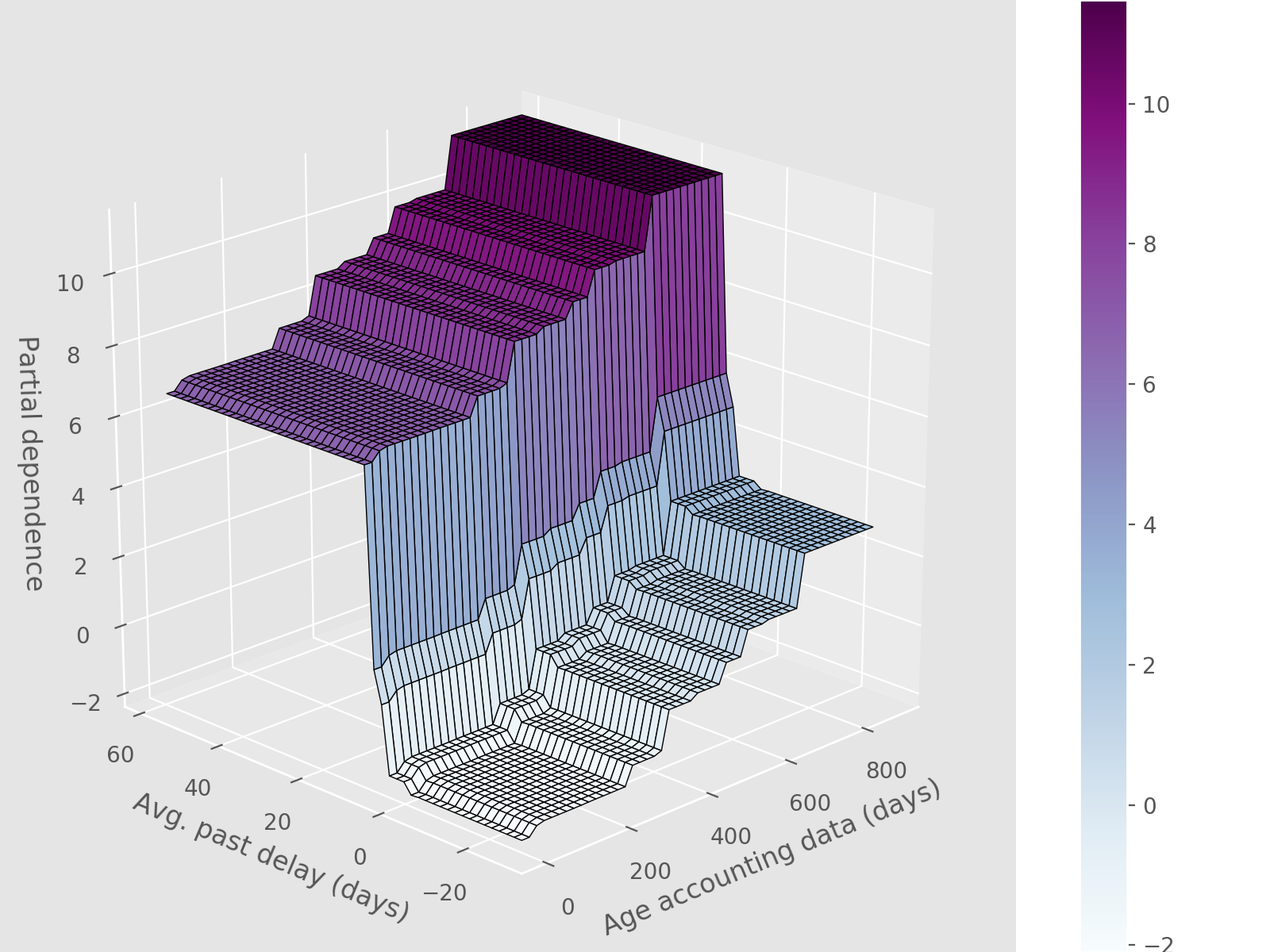}
	\caption{Partial dependence plots illustrating two second-order effects.}
	\label{fig:imp2} 
\end{figure}

In Figure \ref{fig:imp1}, we show partial dependence plots illustrating the main effects of four selected variables with high variable importance. The four variables are a rating from a rating agency (the higher the better is the creditworthiness of a company), a repayment score calculated by Advanon (the higher the better is the repayment history of a company), the age of the accounting data provided by a company, as well as the average number of days of delay by which a company did repay its past loans. The results from this are in line with expectations about default probabilities: companies with higher ratings have a lower default probability. Similarly, a better repayment history and the fact that companies provide more recent accounting data is related to a lower default probability. As can be seen in the plots, three of the four effects are clearly nonlinear. In Figure \ref{fig:imp2}, we additionally show two examples of two-dimensional partial dependence plots illustrating second-order effects.

\subsubsection{Explaining predictions: local partial dependence plots}
Variable importances and partial dependence plots are aggregate measures of the impact of a variable, or a set of variables, on the target variable $Y$. In practice, explaining individual predictions if often at least as important as global model interpretation. The reasons for this include, for instance, that a user of a default prediction model would like to understand how a certain prediction is made in order to make a final decision, or a company can be required by law to be able to provide an explanation. 

In the following, we propose an approach that allows for explaining a specific prediction $$\hat y=\widehat{F}(x'),$$ where $x'$ denotes the values of the predictor variables for which a prediction is made. The approach, which we denote as `local partial dependence plot', is similar to the above partial dependence plot. Instead of averaging over all observations as in \eqref{partdepplot}, one plots $\widehat{F}(X_s,x'_{-s})$ versus $X_s$, where $X_s$ varies in the range of the training data and $x'_{-s}$ contains all other variables. Such a local partial dependence plot allows for answering the question of what would happen to the predicted value if the variable $X_s$ is set to a different value while holding all other variables constant.

In addition, it is often desirable to know which features have the largest influence on a certain prediction. A heuristic importance measure can be obtained by calculating the difference between the largest and smallest value of $\widehat{F}(X_s,x'_{-s})$ where $X_s$ varies in an interval $I_s$. Alternatively, a winsorized version can also be used by calculating the difference between the 97.5\% and 2.5\% quantile of $\widehat{F}(X_s,x'_{-s}), X_s\in I_s$. Concerning the borders of the interval $I_s$, one can use (i) the minimum and maximum of the past data, (ii) a winsorized version of it by excluding, e.g., the lowest and highest 2.5\% of the values, (iii) $x'_{s}\pm \hat\sigma_{x_s}$, where $\hat\sigma_{x_s}$ denotes the empirical standard deviation of $X_s$, or (iv) $x'_{s}\pm \delta_{s}$ where $\delta_{s}$ is a small, local shift around $x'_{s}$. In general, we believe that for tree-based methods, the latter local approach is not ideal since a tree ensemble can locally be very non-smooth, for instance, if an outlier is accounted for. In such situations, local effects can thus be highly noisy.

\begin{figure}[ht!]
	\centering
	\includegraphics[width=0.9\textwidth]{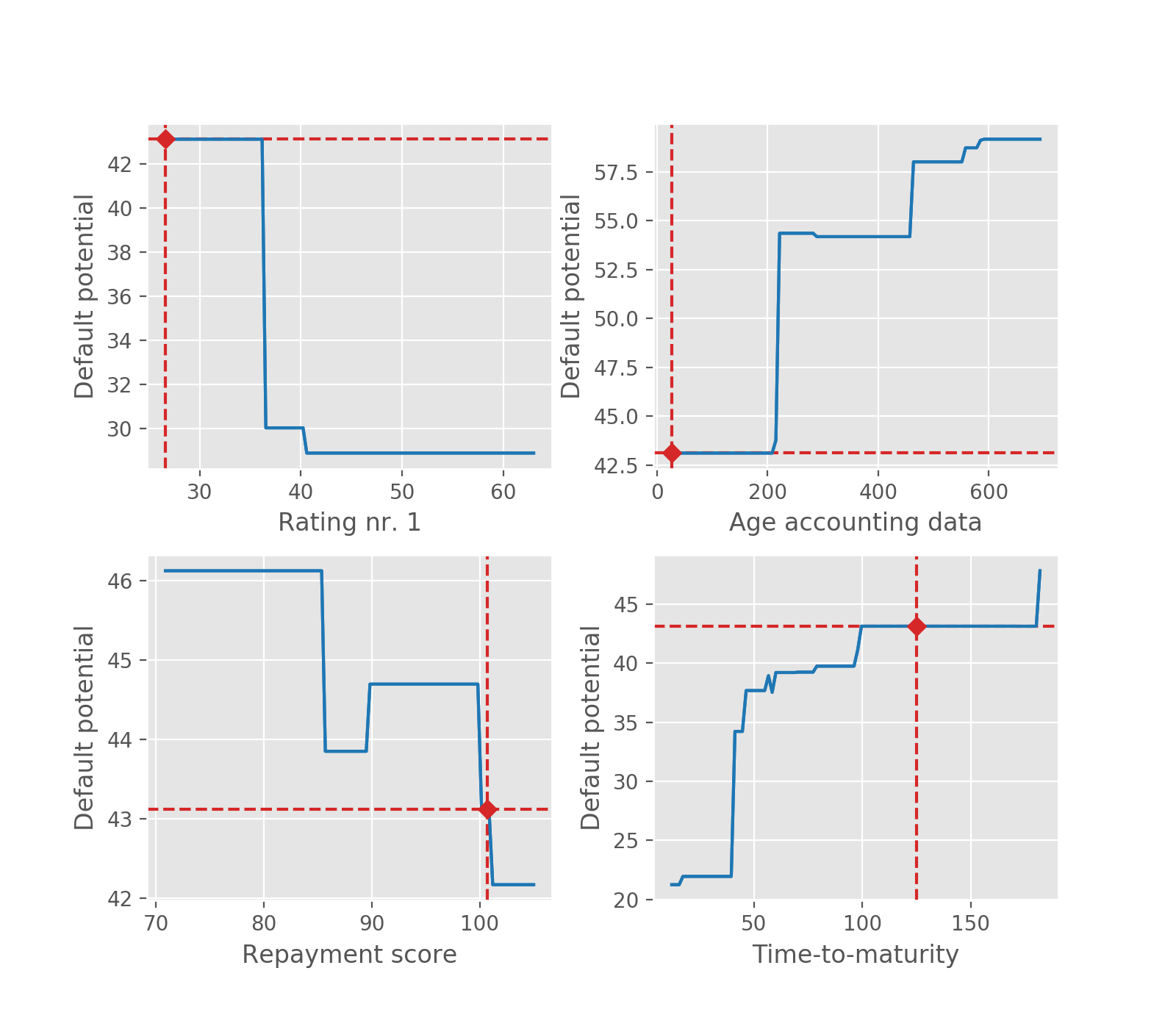}
	\caption{Local partial dependence plot for four variables with high variable importance for one prediction. The red dots and dashed lines represent the predicted value.}
	\label{fig:imp_loc} 
\end{figure}

In Figure \ref{fig:imp_loc}, we show a local partial dependence plot for four variables with high variable importance for one prediction. The red dots and dashed lines represent the predicted value. On the vertical is the predicted default potential (the lower the better). Concerning the rating, the age of the accounting data, and the repayment score, we find similar relationships as in the global partial dependence plots. Concerning the time-to-maturity, the plot shows that the algorithm would predict a lower default potential for this specific loan if the time-to-maturity were lower.

\section{Conclusions}\label{concl}
The Grabit model is a flexible, nonlinear censored regression model that can be applied to various modeling tasks. In particular, this includes binary classification in situations where there is class imbalance but auxiliary data, which is related to the classification mechanism, is available. We have shown in a simulation study and in our default prediction application that the Grabit model can provide substantial and significant gains in predictive accuracy, in particular for small sample sizes. If the decision function is sufficiently complex containing, e.g., nonlinearities, interactions, or a large number of predictor variables, the Grabit model can also provide increased predictive accuracy for datasets of moderate or large size. 

The Grabit model can be extended in several ways. For instance, one can relax the assumption of a constant variance by also relating the variance of the latent variable in Equation \eqref{tobitmod1} to an ensemble of regression trees as it is done for the mean. Further, instead of using functional gradient descent, one can use the functional Newton method in the boosting update for finding both the partition and the leaf values of the trees, see e.g. \citet{sigrist2018gradient}. In addition to trees, one can also use other base learners such as splines \citep{buhlmann2003boosting,hothorn2010model} or reproducing kernel Hilbert space (RKHS) regression functions \citep{sigrist2019KTBoost}. Furthermore, the normality assumption for the latent variable can be relaxed by using another density that is differentiable in the parameter which is related to a tree ensemble.

\section*{Acknowledgments}
We gratefully acknowledge funding by Innosuisse under grant number '25746.1 PFES-ES'. Further, we thank the associate editor as well as two anonymous reviewers for constructive comments and suggestions that helped to improve the paper. We also would like to thank Benjamin Christoffersen for interesting discussions and Dirk Tasche for feedback on the article.

\bibliography{bibfile_grabit}

\begin{appendices}
	\section{Additional results for the Tobit loss}\label{Tdensity}
	It it easily seen that the distribution of $Y$ in the Tobit model in Equation \eqref{tobitmod1} is a mixture of a discrete and a continuous distribution. We denote by $\delta_{a}(y)$, $a\in \mathbb{R}$, the Dirac measure which is the probability measure of a discrete random variable which equals $a$ with probability one. The density of $Y$ with respect to the sum of the Lebesgue measure and the Dirac measures $\delta_{y_l}(y)$ and $\delta_{y_u}(y)$ is then given by
	\begin{equation}\label{tobitdensity}
	\begin{split}
	f_{F(x),\sigma}(y)=&\Phi\left(\frac{y_l-F(x)}{\sigma}\right)\mathbbm{1}_{y_l}(y)+\frac{1}{\sigma}\phi\left(\frac{y-F(x)}{\sigma}\right)\mathbbm{1}_{\{y_l<y<y_u\}}\\&+\left(1-\Phi\left(\frac{y_u-F(x)}{\sigma}\right)\right)\mathbbm{1}_{y_u}(y),
	\end{split}
	\end{equation}
	where $\mathbbm{1}_{A}(y)$, $A\subset \mathbb{R}$, denotes the indicator function which equals one if $y\in A$ and zero otherwise, $\phi(x)$ and $\Phi(x)$ are the standard normal density and cumulative distribution. 
	
	The negative log-likelihood of the Tobit model, which is the loss $L(y,F(x))$ used in the Grabit model, is then given by
	\begin{equation}\label{tobitloss}
	\begin{split}L(y,F(x))=&-\log\left(\Phi\left(\frac{y_l-F(x)}{\sigma}\right)\right)\mathbbm{1}_{y_l}(y)\\
	&+\left(\frac{(y-F(x))^2}{2\sigma^2}+\log(\sigma)+0.5\log(2\pi)\right)\mathbbm{1}_{\{y_l<y<y_u\}}\\
	&-\log\left(1-\Phi\left(\frac{y_u-F(x)}{\sigma}\right)\right)\mathbbm{1}_{y_u}(y).
	\end{split}
	\end{equation}
	As mentioned in the main text, the Tobit loss is asymmetric in $F(x)$ for the cases $y=y_l$ and $y_u$.
	
	The gradient $\frac{\partial L(y,F)}{\partial F}$ and the second derivative $\frac{\partial^2 L(y,F)}{\partial^2 F}$ of the Tobit loss are given by:
	\begin{equation}\label{gradient}
	\begin{split}\frac{\partial L(y,F)}{\partial F}=&\frac{\phi\left(\frac{y_l-F(x)}{\sigma}\right)}{\sigma\Phi\left(\frac{y_l-F(x)}{\sigma}\right)}\mathbbm{1}_{y_l}(y)-\frac{y-F(x)}{\sigma^2}\cdot\mathbbm{1}_{\{y_l<y<y_u\}}\\
	&-\frac{\phi\left(\frac{y_u-F(x)}{\sigma}\right)}{\sigma\left(1-\Phi\left(\frac{y_u-F(x)}{\sigma}\right)\right)}\mathbbm{1}_{y_u}(y)
	\end{split}
	\end{equation}
	and
	\begin{equation}\label{secderiv}
	\begin{split}\frac{\partial^2 L(y,F)}{\partial^2 F}=&\frac{\phi\left(\frac{y_l-F(x)}{\sigma}\right)}{\sigma^2\Phi^2\left(\frac{y_l-F(x)}{\sigma}\right)}\left(\frac{y_l-F(x)}{\sigma}\Phi\left(\frac{y_l-F(x)}{\sigma}\right)+\phi\left(\frac{y_l-F(x)}{\sigma}\right)\right)\mathbbm{1}_{y_l}(y)\\
	&+\frac{1}{\sigma^2}\mathbbm{1}_{\{y_l<y<y_u\}}\\
	&-\frac{\phi\left(\frac{y_u-F(x)}{\sigma}\right)}{\sigma^2\left(1-\Phi\left(\frac{y_u-F(x)}{\sigma}\right)\right)^2}\left(\frac{y_u-F(x)}{\sigma}\left(1-\Phi\left(\frac{y_u-F(x)}{\sigma}\right)\right)-\phi\left(\frac{y_u-F(x)}{\sigma}\right)\right)\mathbbm{1}_{y_u}(y).
	\end{split}
	\end{equation}
	
	
	
	\section{Classification tree and random forest}\label{tree_rf}
	For the classification tree used in Section \ref{defaulpred}, we adopt the approach of \citet{breiman1984classification} with the Gini impurity measure as the splitting criterion. Similarly to boosting, a random forest consists of an ensemble of classification trees. In contrast to boosting, the trees are built in parallel and not in a sequential manner, which means that there is less dependence between the trees. The idea is that one single deep tree has high variance and low bias, and when aggregating such trees, one can reduce the variance while still having a low bias. Different trees are obtained using subsampling with replacement of the data, a process denoted by bagging. Additional independence between trees is created by only considering a subset of the variables instead of all variables when making splits in the tree algorithm. See \citet{friedman2001elements} for more details on classification trees and random forests.
	
	\section{Choice of tuning parameters}\label{tune_pars}
	In the following, we describe how the tuning parameters of the models used in the application in Section \ref{defaulpred} are chosen. Below, we list for each tuning parameter of each model, the candidate choices of the tuning parameter. For each model, we then consider the full grid obtained when using all combinations of individual parameters. We use the temporal cross-validation scheme described in Section \ref{eval_appl} with the area under the receiver operating characteristic (AUROC) as a measure of fit. Specifically, for each observation and each parameter combination, we fit a model based on past data, where we require that there at least 100 past data points. We then compare the thus obtained predictions with the observed data using the AUROC.
	
	We consider the following tuning parameters for the tree-boosted methods (boosted Logit, boosted multiclass Logit and Grabit): the number of trees $M$, the learning rate $\nu$, and the depth of the trees $T$. These are chosen among the following combinations: $M\in\{10,100,1000\}$, $\nu\in\{0.1,0.01,0.001\}$, and $T\in\{3,5,10\}$. For the Grabit model, we additionally select $\sigma$ from $\{0.01,0.1,1,10,100\}$. For the random forest algorithm, we consider the following tuning parameters: the number of trees $M\in\{10,100,1000\}$, the fraction of the number of variables considered for making a split when growing trees $\rho\in \{50\%, 75\%,100\%\}$, and the tree depth $T\in\{3,5,10\}$. For the tree, we consider the minimum number of samples per leaf $S\in\{1,10,100\}$ and the depth of the tree $T\in\{3,5,10,\infty\}$, where $\infty$ means that the tree is grown until all leaves are pure or until all leaves contain less than the minimum number of samples per leaf. For the neural network, we use two hidden layers with five and two nodes, respectively, and rectified linear units (ReLU) as activation functions. No attempt is made to search for an optimal network structure of the neural network. As we mention below in the text, the sample size is too small to have an additional validation set, which we could use for choosing the potentially many tuning parameters of a neural network (e.g. number of hidden layers, number of nodes per layer, different activation functions, several regularization options) without resulting in severe in-sample overfitting.
	
	We obtain the following combination of tuning parameters. Boosted Logit: $\nu=0.1$, $M=1000$, $T=3$. Boosted multiclass Logit: $\nu=0.01$, $M=1000$, $T=5$. Grabit: $\nu=0.1$, $M=100$, $T=3$, $\sigma=1$. Random forest: $M=1000$, $T=10$, $\rho=1$. Classification tree: $S=1$ and $T=3$.

\begin{figure}[ht!]
		\centering
		\includegraphics[width=0.9\textwidth]{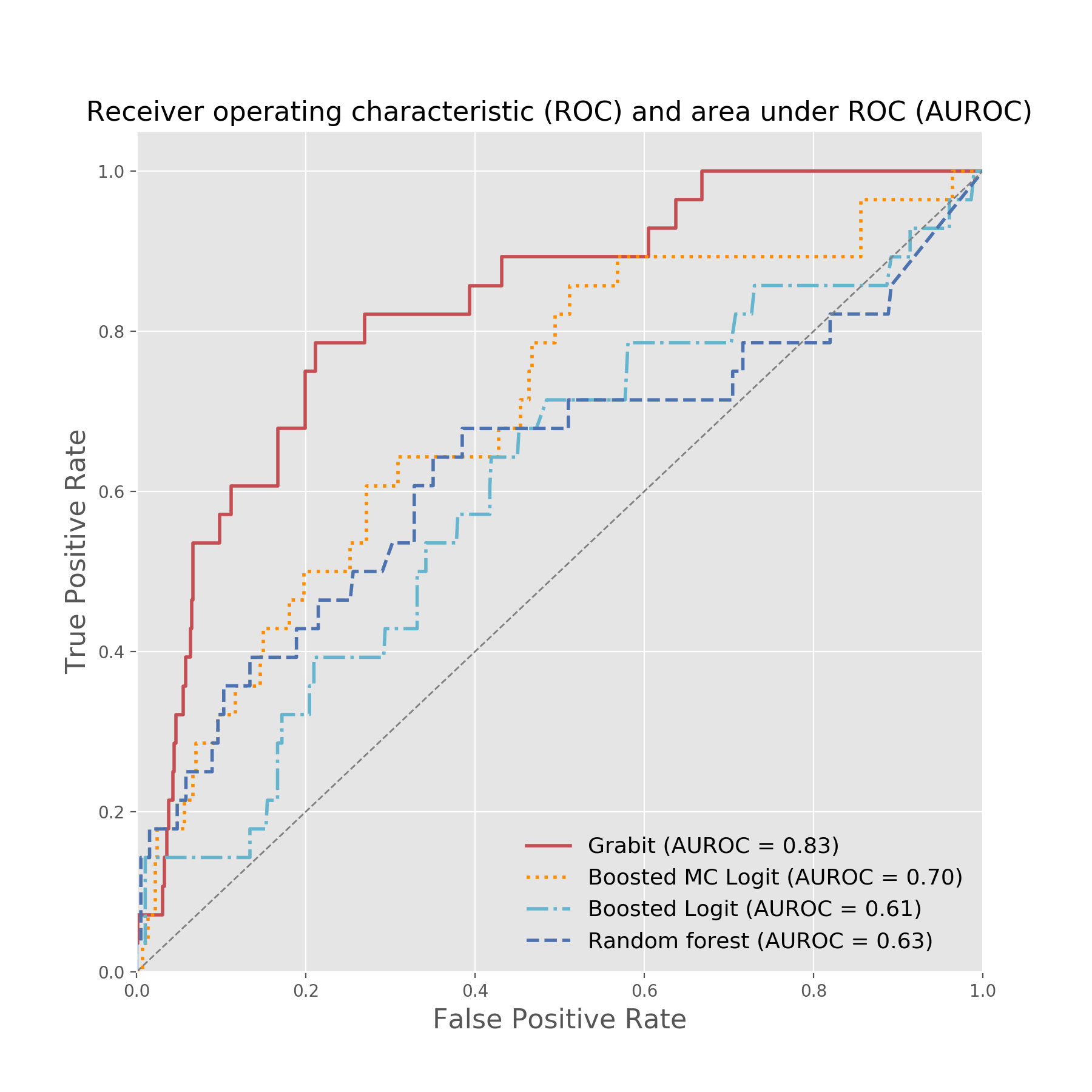}
		\caption{Comparison of different models using receiver operating characteristic (ROC) and area under ROC (AUROC) when using non-optimal tuning parameters for the tree-boosted methods (boosted Logit, boosted multiclass Logit, Grabit) and the random forest.}
		\label{fig:roc_non_opt} 
	\end{figure}

	Further, the following combination of tuning parameters results in the second best predictive accuracy. Boosted Logit: $\nu=0.1$, $M=10$, $T=3$. Boosted multiclass Logit: $\nu=0.1$, $M=1000$, $T=5$. Grabit: $\nu=0.01$, $M=1000$, $T=3$, $\sigma=1$. Random forest: $M=1000$, $T=5$, $\rho=0.5$. Figure \ref{fig:roc_non_opt} shows the results for the second best combination of tuning parameters.
	
\end{appendices}

\end{document}